\documentclass[twocolumn]{aastex63}
\usepackage{amsmath,bm}
\usepackage{xcolor}
\usepackage[english]{babel}

\newcommand{\kms}{$\mathrm{km\,s^{-1}}$}
\newcommand{\msun}{M_{\sun}}
\newcommand{\msunh}{h^{-1}M_{\sun}}
\newcommand{\msunhh}{h^{-2}M_{\sun}}
\newcommand{\hi}{H{\sc i}}

\shorttitle{\hi-halo Mass Relation}
\shortauthors{H. Guo et al.}

\begin{document}

\title{Direct Measurement of the \hi--Halo Mass Relation through Stacking}

\author[0000-0003-4936-8247]{Hong Guo}
\affiliation{Key Laboratory for Research in Galaxies and Cosmology, Shanghai Astronomical Observatory, Shanghai 200030, China}
\altaffiliation{guohong@shao.ac.cn}

\author[0000-0002-5434-4904]{Michael G. Jones}
\affiliation{Instituto de Astrof\'isica de Andaluc\'ia, Glorieta de la Astronom\'ia s/n, 18008 Granada, Spain}

\author[0000-0001-5334-5166]{Martha P. Haynes}
\affiliation{Cornell Center for Astrophysics and Planetary Science, Space Sciences Building, Cornell University, Ithaca, NY 14853, USA}

\author{Jian Fu}
\affiliation{Key Laboratory for Research in Galaxies and Cosmology, Shanghai Astronomical Observatory, Shanghai 200030, China}

\begin{abstract}
We present accurate measurements of the total \hi\ mass in dark matter halos of different masses at $z\sim0$, by stacking the \hi\ spectra of entire groups from the Arecibo Fast Legacy ALFA Survey. The halos are selected from the optical galaxy group catalog constructed from the Sloan Digital Sky Survey DR7 Main Galaxy sample, with reliable measurements of halo mass and halo membership. We find that the \hi-halo mass relation is not a simple monotonic function, as assumed in several theoretical models. In addition to the dependence of halo mass, the total \hi\ gas mass shows strong dependence on the halo richness, with larger \hi\ masses in groups with more members at fixed halo masses. Moreover, halos with at least three member galaxies in the group catalog have a sharp decrease of the \hi\ mass, potentially caused by the virial halo shock-heating and the AGN feedback. The dominant contribution of the \hi\ gas comes from the central galaxies for halos of $M_{\rm h}<10^{12.5}\msunh$, while the satellite galaxies dominate over more massive halos. Our measurements are consistent with a three-phase formation scenario of the \hi-rich galaxies. The smooth cold gas accretion is driving the \hi\ mass growth in halos of $M_{\rm h}<10^{11.8}\msunh$, with late-forming halos having more \hi\ accreted. The virial halo shock-heating and AGN feedback will take effect to reduce the \hi\ supply in halos of $10^{11.8}\msunh<M_{\rm h}<10^{13}\msunh$. The \hi\ mass in halos more massive than $10^{13}\msunh$ generally grows by mergers, with the dependence on halo richness becoming much weaker.   
\end{abstract}


\section{Introduction}
How galaxies obtain their gas, form stars and quench, is the very basic question in understanding galaxy formation and evolution. In the standard paradigm of galaxy formation \citep[e.g.,][]{Rees+1977, Silk+1977, White+1978}, gas infalling into a dark matter halo suffers from virial shock-heating around the halo virial radius, which impedes the efficiency of gas cooling. Detailed studies in simulations suggest that the virial shocks are only important above a critical shock-heating halo mass $\sim10^{12}\msun$ \citep[e.g.,][]{Birnboim+2003, Keres+2005, Dekel+2006}. While the feedback from active galactic nuclei (AGN) is also thought to be important around the similar halo mass scale \citep[e.g.,][]{DiMatteo+2005}, it is then crucial to constrain the strength of AGN feedback from the relation between the cold gas mass and its host dark matter halo mass.

However, it is still observationally challenging to directly probe this relation. The majority of the cold gas in the universe is comprised of the neutral hydrogen. While the amount of atomic neutral hydrogen (\hi) can be reliably measured at $z\sim0$ through the 21 cm hyperfine emission line, the molecular neutral hydrogen ($\rm{H_2}$) is generally difficult to probe directly. In the past decade, there have been lots of efforts to map the \hi\ distribution in the universe, e.g., the \hi\ Parkes All-Sky Survey \citep[HIPASS;][]{Barnes+2001,Meyer+2004}, the Arecibo Fast Legacy ALFA Survey \citep[ALFALFA;][]{Giovanelli+2005}, and the GALEX Arecibo SDSS Survey \citep[GASS;][]{Catinella+2010}. There are only a few surveys to measure the $\rm{H_2}$ content using the tracer of CO emission lines, e.g., the COLD GASS survey \citep{Saintonge+2011}, the Bima survey of nearby galaxies \citep[BIMA SONG][]{Helfer+2003}, the HERA CO-Line Extragalactic Survey \citep[HERACLES][]{Leroy+2009}, and the JINGLE survey \citep{Saintonge+2018}. As the size and uniformity of existing \hi\ samples far exceed those of $\rm{H_2}$, we focus on \hi\ content in this work. 

Another difficulty of constraining the \hi-halo mass relation is the estimation of the dark matter halo mass. \cite{Guo+2017} measured the spatial clustering of the \hi-selected galaxies in the ALFALFA 70\% sample and constrained the average halo masses for different \hi\ mass samples. More importantly, they found that the distribution of \hi-selected galaxies is not only dependent on the dark matter halo mass, but also related to the formation history of the host dark matter halos. Galaxies that are richer in \hi\ tend to live in halos formed more recently, which is generally referred to as the halo assembly bias effect \citep{Gao+2005}. It complicates the \hi-halo mass relation with the additional dependence on other halo parameters related to the formation history. One such parameter is the halo angular momentum. There is various evidence that the \hi-rich galaxies tend to have higher halo spin parameters \citep[e.g.,][]{Huang+2012,Maddox+2015,Obreschkow+2016,Lutz+2018}. Therefore, it is necessary to include the effect of halo formation history when considering the \hi-halo mass relation.

Since blind \hi\ surveys like ALFALFA have selection effects arising from the \hi\ flux limit, as well as the dependence on the \hi\ line profile width ($W_{50}$) \citep{Haynes+2011}, optically visible galaxies with low \hi\ fluxes and broad line profiles will not be detected in such radio surveys. Estimating halo masses for the ALFALFA-detected \hi\ sources essentially measures the relation of $\langle M_{\rm h}|M_{\rm HI}\rangle$ as in \cite{Guo+2017}, i.e., the average halo mass at a given \hi\ mass. It is significantly different from the relation of $\langle M_{\rm HI}|M_{\rm h}\rangle$, which measures the total \hi\ mass contained in halos of different masses, including those not detected by the observations \citep[see e.g., Fig. 3 of][]{Kim+2017}. While $\langle M_{\rm h}|M_{\rm HI}\rangle$ can only be used to quantify the halo mass for the \hi-rich galaxies detected by radio surveys, $\langle M_{\rm HI}|M_{\rm h}\rangle$ can be directly compared to the theoretical models with various quenching mechanisms that affect the total cold gas mass in different halos \citep[see e.g., the review of][]{Man+2018}. The measurement of $\langle M_{\rm HI}|M_{\rm h}\rangle$ is also important for current and future 21 cm intensity mapping projects, since the 21 cm power spectrum in the linear order is proportional to the product of the total \hi\ bias $b_{\rm HI}$ and the cosmic \hi\ abundance $\Omega_{\rm HI}$ \citep[e.g.,][]{Villaescusa-Navarro+2018,Obuljen+2019,Wolz+2019}.

Although $\langle M_{\rm HI}|M_{\rm h}\rangle$ has been investigated extensively in different theoretical models, e.g., empirical models \citep{Barnes+2014,Paul+2018,Obuljen+2019}, semi-analytical models \citep{Kim+2017,Zoldan+2017,Baugh+2019} and hydrodynamical simulations \citep{Villaescusa-Navarro+2018}, there still lacks direct observational measurement of this \hi-halo mass relation. \cite{Ai+2018} have attempted to quantify the total \hi\ mass for rich galaxy groups with at least eight members using the detected sources in the ALFALFA 70\% sample. They derived the group \hi\ mass fraction by summing up the \hi\ masses for all detected \hi\ sources and correcting for the missing ones based on the scaling relation between the \hi\ mass and those of galaxy luminosity and color. Given the large uncertainties in the \hi\ scaling relation for targets below the observational detection limit, one straightforward solution to properly take into account all the \hi\ emitting sources is to stack the \hi\ signals for entire galaxy groups with reliable halo mass estimates.

The \hi\ spectral stacking technique has previously been extensively applied in quantifying the relationships of \hi\ gas fraction with galaxy properties, such as stellar mass, color, star formation rate and stellar surface density \citep[see e.g.,][]{Verheijen+2007,Fabello+2011,Gereb+2015,Brown+2015,Brown+2017}, as well as in constraining $\Omega_{\rm HI}$ at various redshifts \citep{Lah+2007,Delhaize+2013,Rhee+2013,Hu+2019}. Unlike with the traditional method of stacking the \hi\ spectra of single galaxies, applying \hi\ stacking to dark matter halos offers the great advantage that the stacking results should not be significantly affected by the spatial resolution of the \hi\ data, because the sizes of the dark matter halos are typically much larger than or at least comparable to the beam sizes of the radio telescopes. Therefore the effect of confusion from different halos would be minimal for such an experiment.

In this paper, we will directly measure $\langle M_{\rm HI}|M_{\rm h}\rangle$ by stacking the \hi\ spectra for dark matter halos selected from the galaxy group catalog. Such a stacking method is based on the single \hi\ spectrum from each entire group, which includes the contribution from all the \hi\ gas in the halos. The structure of the paper is as follows. In \S\ref{sec:data}, we describe the galaxy samples. We briefly introduce our \hi\ stacking method in \S\ref{sec:method} and present the results in \S\ref{sec:results}. We summarize and discuss the results in \S\ref{sec:discussion} and \S\ref{sec:summary}. 

Throughout this paper, we assume a spatially flat $\Lambda$CDM
cosmology, with $\Omega_{\rm m}=0.307$, $h=0.678$,
$\Omega_{\rm b}=0.048$ and $\sigma_8=0.823$, consistent with the
constraints from Planck \citep{PlanckCollaboration+2014}.
\section{Data}\label{sec:data}

\subsection{The ALFALFA survey}

The ALFALFA survey \citep{Giovanelli+2005} blindly mapped the \hi \ line emission over approximately 6900 deg$^2$ of the Northern sky in the redshift range $-2000 < cz_\odot/\rm{km\,s^{-1}} < 18000$. The survey utilized a two pass, drift scan strategy, with each second passage offset by half a beam width. This was extremely time efficient and resulted in highly uniform coverage. The survey footprint was split into two regions in the Northern Spring ($07^{\rm{h}}30^{\rm{m}} < \rm{RA} < 16^{\rm{h}}30^{\rm{m}}$) and Fall ($22^{\rm{h}} < \rm{RA} < 03^{\rm{h}}$) skies in the Declination range $0^\circ < \rm{Dec} < 36^\circ$. The final source catalog \citep{Haynes+2018} contains over 30000 extragalactic sources. 

In this work we focus exclusively on the Spring sky portion of the survey, as this is where there is appreciable overlap with the footprint of the Sloan Digital Sky Survey (SDSS) legacy spectroscopic survey \citep{York+2000}. In the data processing stage, the survey area of ALFALFA is split into a pre-defined set of grids, with each grid square being 2.4$^\circ$ on a side and spaced approximately 2$^\circ$ apart. Each spatial grid is also divided into four overlapping ranges in heliocentric velocity, to make four spectral cubes for each spatial grid square \citep[for further details refer to][]{Haynes+2011,Haynes+2018}. Each spectral cube has dimensions of $144 \times 144 \times 1024$ corresponding to a pixel angular size of 1\arcmin, or approximately a quarter of the beam diameter ($3.8\arcmin \times 3.3\arcmin$). In addition to maps of the \hi \ line flux density, each cube contains a normalized weight map which indicates how much of the input data have been flagged for poor quality or radio frequency interference (RFI) at any given point in the cube. The typical rms noise of the data is approximately 2 mJy per 5 \kms \ channel, although the data are Hanning smoothed to a spectral resolution of about 10 \kms.

\subsection{Galaxy Group catalog}
In order to stack the \hi\ signals for galaxy groups in the SDSS region, we use the SDSS galaxy group catalog from \cite{Lim+2017}, which is an extension to the early SDSS group catalog of \cite{Yang+2007}. This group catalog is based on the SDSS DR7 Main Galaxy Sample \citep{Albareti+2017}, but it incorporates the redshifts for the missing galaxies due to fiber collisions from different sources (e.g. the later Data Release 13 and other surveys). We refer the readers to \cite{Lim+2017} for more details. We adopt their SDSS group catalog with all galaxies having spectroscopic redshifts, which is about 98\% complete compared to the full target sample. The halo masses in this group catalog are estimated using the proxy of galaxy stellar mass. The halo radius $r_{200}$ is estimated from the definition that the mean mass density within $r_{200}$ is 200 times the mean density of the universe at the given redshift, i.e., 
\begin{equation}
M_{\rm h}=200\bar{\rho}_{\rm m}(1+z)^3\frac{4\pi}{3} r_{200}^3
\end{equation}
where $\bar{\rho}_{\rm m}$ is the mean background density of the universe at $z=0$.

The halo mass estimates in \cite{Lim+2017} have been demonstrated to be unbiased using mock catalogs. The typical scatter is less than $0.2$\,dex. As shown in their Figures 7 and 8, halos with $\log(M_{\rm h}/\msun)>11.5$ are all complete at $z<0.05$ within the SDSS. As will be discussed in Section~\ref{subsubsec:comp}, although the SDSS DR7 galaxy catalog is a flux-limited sample, galaxies in the observed halos are basically complete above the stellar mass of $M_{\ast}>10^{9.5}\msunhh$, which means that the galaxy group catalog is missing some low-mass galaxies. This has two direct implications. Firstly, it emphasizes the importance of stacking the total \hi\ signal for each halo, rather than stacking the \hi\ spectra for the observed halo member galaxies, which will be biased towards the gas-rich massive galaxies. As the SDSS target sample selection is based on the galaxy luminosity, we do not expect any strong selection bias when we stack halos in different mass bins. So even though halos are not complete for $\log(M_{\rm h}/\msun)<11.5$, the stacking measurements do not suffer from the selection effects. Secondly, the richness information for each halo in the group catalog, i.e., the number of member galaxies, is associated with certain stellar mass thresholds. We find that the halo richness is reliable for galaxies with $M_{\ast}>10^7\msunhh$.

For the purpose of matching the ALFALFA survey depth, we limit the redshift range of the halos in the group catalog to be $0.0025<z<0.06$, and we only use the group galaxies in the ALFALFA Spring sky. The final sample includes $28,910$ groups and $53,653$ galaxies \footnote{We note that the group catalog includes many groups with a single member, i.e., halo richness equal to one.}. By cross-matching with the ALFALFA final source catalog, we find that only $15,211$ galaxies have measured \hi\ masses, i.e. the majority of the galaxies in the group catalog are below the ALFALFA detection limit \citep{Haynes+2011}. It further emphasizes the importance of using the \hi\ signal stacking method to measure reliably the average \hi\ mass in each halo mass bin \citep{Jones+2020}. 

\section{Stacking Method}\label{sec:method}

\begin{figure}
    \centering
    \includegraphics[width=\columnwidth]{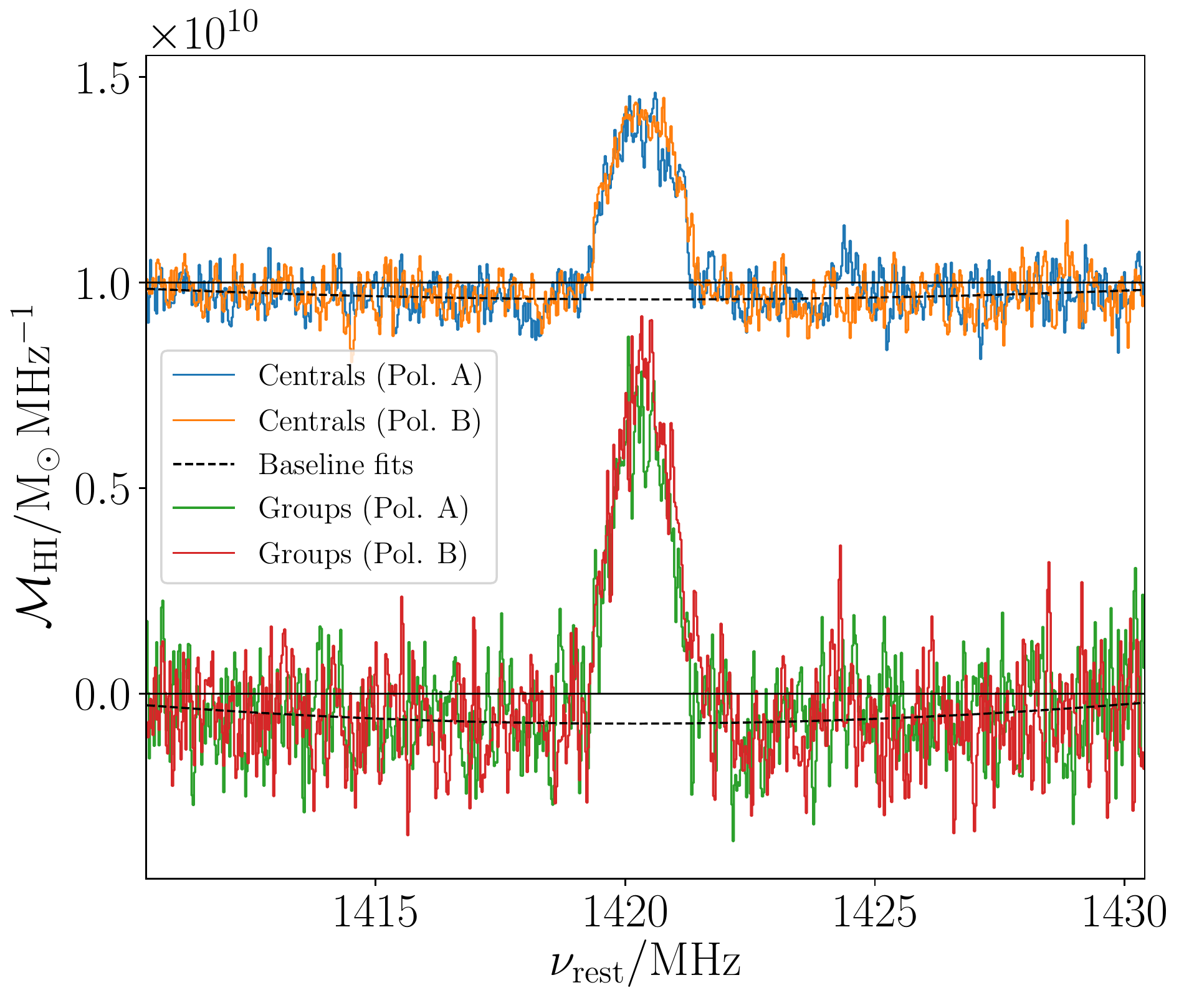}
    \caption{Example stacked spectra for both centrals and groups before the re-baselining was performed. These data are for the $12.25 < \log (M_\mathrm{halo}/h^{-1}\,\mathrm{M_\odot}) < 12.5$ bin and for groups with at least 3 members. The centrals-only stack is vertically offset to show both spectra in the same figure. The two linear polarizations are shown for each spectrum with different colors and the 2nd order polynomial baseline fits are shown with dashed black lines.}
    \label{fig:example_stack}
\end{figure}

In this work we use the ALFALFA \texttt{IDL} stacking software developed by \citet{Fabello+2011} to extract spectra of each group and central in the group catalog. This software spatially integrates over a square aperture and returns a 1-dimensional spectrum of the \hi \ flux density. The velocity range covered corresponds to the entire range of the ALFALFA spectral cube with the nearest centre frequency to the expected frequency of the \hi \ emission of the target object, given its redshift. In addition to the flux density, a weight spectrum, integrated over the same sky aperture, is also generated, allowing us to track the impact of missing or poor quality data or contamination by RFI.

While \citet{Fabello+2011} used a fixed aperture size for all stacked galaxies, the groups are frequently considerably larger than the single ALFA beam width. Therefore, each of our apertures is tailored to each group or central. We use the $r_{200}$ values to define the angular size of the aperture ($2r_{200}$) for each group, rounded up to a whole number of arcmin (pixels). The apertures for the centrals were all conservatively chosen to be 200 kpc in diameter. The latest measurement of the \hi\ size--mass relation shows that the largest \hi\ discs are between 100 and 200 kpc in diameter \citep[see e.g., Figure~1 of][]{Wang+2016}. Thus, this choice of aperture ensures that \hi\ flux from a target central galaxy will not be missed. The Hubble distance for each group was used to convert this to an angular size, which was again rounded up to an integer number of arcmin. This choice of aperture for the centrals undoubtedly leads to considerable contributions from confused emission, which we discuss further in the following sections. In both cases, we set a minimum aperture diameter of 8\arcmin \ (approximately two beam widths). This means that for groups with distances greater than $\sim$85 Mpc, the aperture for the central will be larger than 200 kpc, likely leading to additional confusion.

In order to avoid re-gridding the ALFALFA data, any groups (and their centrals) which overlapped a boundary of the cube that contained them (specifically the one with the nearest center position on the sky) were discarded. This resulted in the removal of approximately 2\% of the groups (centrals). The four spectral cubes that each ALFALFA grid square is divided into overlap with the neighboring cube by $\sim$700 \kms, while the  velocity dispersion estimated by \citet{Lim+2017} for a group of $\log ( M_{\rm{halo}} h^{-1} / M_{\odot} ) = 14$ is 418 \kms. Therefore, the velocity axis was ignored when making these cuts.

The spectrum extraction process was performed for every group and central available. Of the total groups and centrals in the catalog, 25906 group spectra (90\%) and 25868 (89\%) central spectra were successfully extracted from the ALFALFA cubes. The targets which were not extracted were discarded due to the low weight of their spectra or proximity to grid boundary as described in the previous paragraph. The spectral extraction software of \citet{Fabello+2011} discards spectra if more than 40\% of the channels have a weight value of less than 0.5. In practice these spectra would mainly contribute noise to any stacks because too much of their data has been flagged for RFI or is missing coverage. The successfully extracted spectra were then divided into halo mass bins and separate stacks produced for each mass bin. This will be described further in the next section. The following paragraphs describe how the extracted spectra were stacked in a general sense.

Once the spectra were extracted, we combined them using our own \texttt{Python} script. Each spectrum is shifted such that the expected frequency of its \hi \ emission falls in the central channel of the stack spectrum. This is done to the nearest channel ($\sim$5 \kms) so that the spectra do not need to be re-binned. The individual spectra are then converted from units of mJy to $\rm{M_\odot \, MHz^{-1}}$ following equation 45 of \citet{Meyer+2017}. Any regions of the spectra which have a normalized weight value of less than 50\% are discarded, as these regions can contain residual unflagged bright RFI. We also discard any spectra with spikes that exceed 100 times the expected rms noise, as these would contribute a lot of noise to the final stack. In total, this results in 24443 groups which contribute to the final stacks. The spectra are co-added with each weighted by $1/\sigma_{\rm{rms}}^{2}$. Here the rms noise in mJy (not $\rm{M_\odot \, MHz^{-1}}$) is used to avoid the weighting being distance dependent. Throughout this process the two polarizations recorded in ALFALFA were treated entirely independently, which provided an additional means to verify that there was no polarized interference affecting the final stacks.

To measure the \hi \ mass in each stack, the stack spectra need to first be re-baselined. We observed that the continuum level in the stacks was generally slightly negative, probably because the original ALFALFA baselining procedure fit within regions containing line emission, but at extremely low SNR (i.e. SNR $<$ 1), such that it can only be perceived in stacks of hundreds of targets. The central 20 MHz of the stack spectrum (with the inner central 5 MHz excluded) was used to fit and remove a 2nd order baseline (Figure~\ref{fig:example_stack}). 

From this point onward, stacks of centrals and stacks of groups were treated differently. The peak of stacked \hi \ emission in a group stack was fit with a Gaussian profile and the flux (measured, not the fit) within $\pm 3\sigma$ was summed to estimate the average \hi \ mass of the groups in the stack. For the stacks of centrals we did not fit a Gaussian to the profiles because, in most cases, considerable confusion was likely. Instead we summed all emission within a $\pm$300 \kms \ window. Very few galaxies have \hi \ line widths greater than 600 \kms \ \citep[e.g][]{Papastergis+2011}, meaning this window will contain all the targeted line emission, except in extreme cases.

The final step of the stacking process was to estimate the uncertainties in the stack masses. This was done through bootstrapping the entire stacking process. For each final stack, 1000 iterations were generated where the input catalogue of targets was randomly sampled (with replacement) to construct a bootstrap sample of the same size. The uncertainty in the mass measurements was taken as the standard deviation of these 1000 iterations.

\section{Results}\label{sec:results}
\subsection{\hi-Halo Mass Relation}
\begin{figure*}
	\centering
	\includegraphics[width=0.7\textwidth]{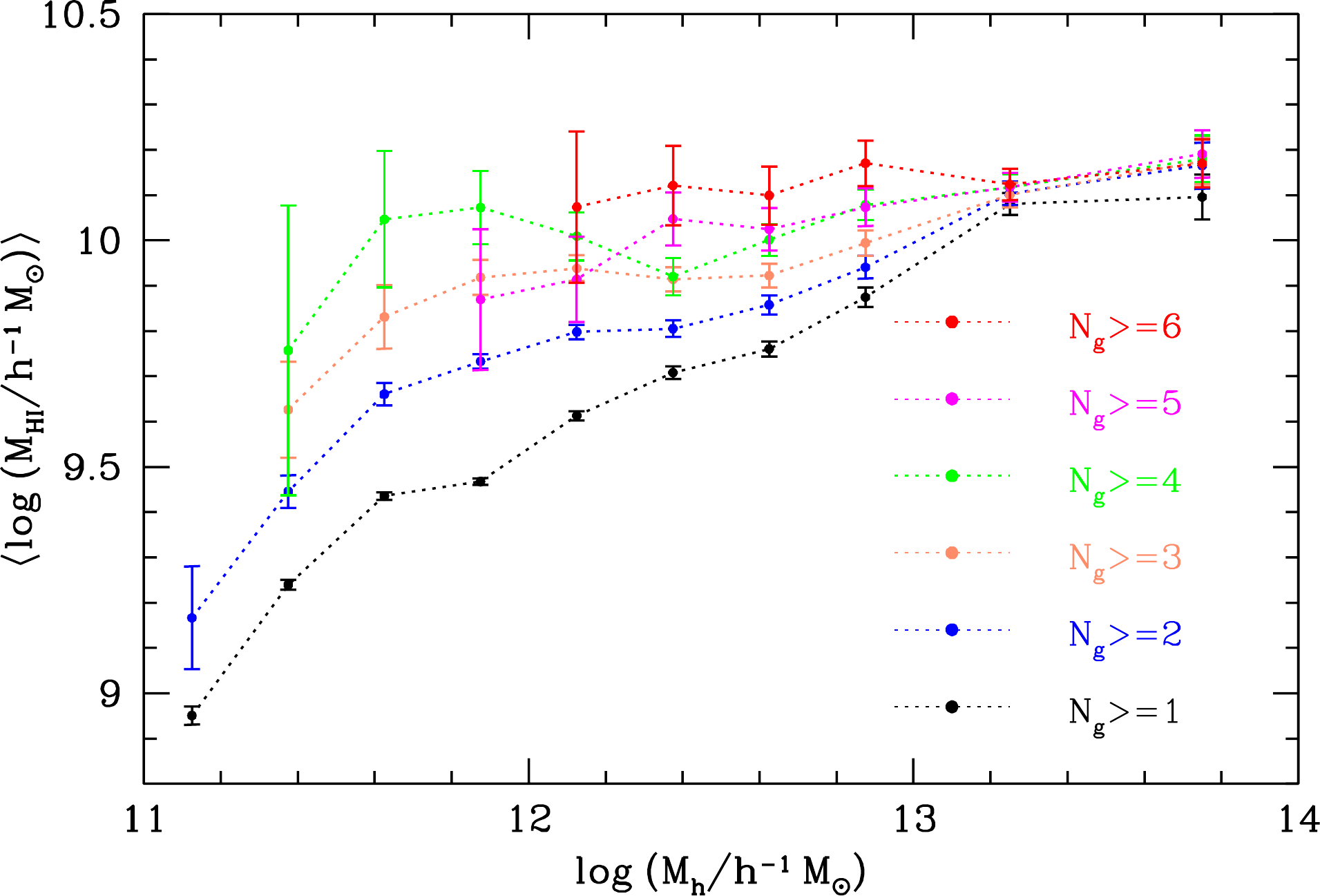}
	\caption{Measurements of total \hi\ masses for halos in different mass bins. We show the measurements for different richness groups in different colors as labeled. There is an increasing turn-over feature with larger group richness around the halo mass scale $\sim10^{12}\msun$.}
	\label{fig:hihm}
\end{figure*}
We show in Figure~\ref{fig:hihm} the average \hi-halo mass relation, i.e., $\langle M_{\rm HI}|M_{\rm h}\rangle$, from our \hi\ spectral stacking technique. The points with the black dotted line are the measurements using all the halos for which we extracted spectra. We also present the measurements for halos of different richness $N_{\rm g}$, i.e. number of member galaxies in the group catalog, shown as points and dotted lines with different colors. We find that there is a clear dependence of the \hi-halo mass relation on the halo richness. Halos with higher richness (at fixed halo mass) generally tend to have larger average \hi\ masses. 

We note that the total \hi\ mass in the halo is not a simple monotonically increasing function of the halo mass. The total \hi\ mass in halos of $N_{\rm g}\ge 2$ tends to increase to a plateau around $M_{\rm h}\sim10^{12.4}\msunh$ and  then sharply increase again. For halos with richness $N_{\rm g}\ge3$ and $N_{\rm g}\ge4$, there is a pronounced bump feature in the \hi-halo mass relation for $M_{\rm h}<10^{12.5}\msunh$ with the peak at around $M_{\rm h}\sim10^{11.9}\msunh$. For halos of higher richness of $N_{\rm g}\ge5$, we are not able to clearly detect such a feature, due to the large errors in the measurements and lack of low-mass halos with a high richness.

For halos with richness of $N_{\rm g}\ge6$, $\langle\log(M_{\rm HI}/\msunh)\rangle$ approaches a constant of $10.1$, without any strong dependence on the halo mass. We have stacked halos of higher richness and find the same behavior. It provides a direct estimate of the total \hi\ mass for rich galaxy clusters with different halo masses. At the massive end, more and more halos have multiple member galaxies and the dependence on the halo richness seems to become much weaker.

\begin{figure*}
	\centering
	\includegraphics[width=0.8\textwidth]{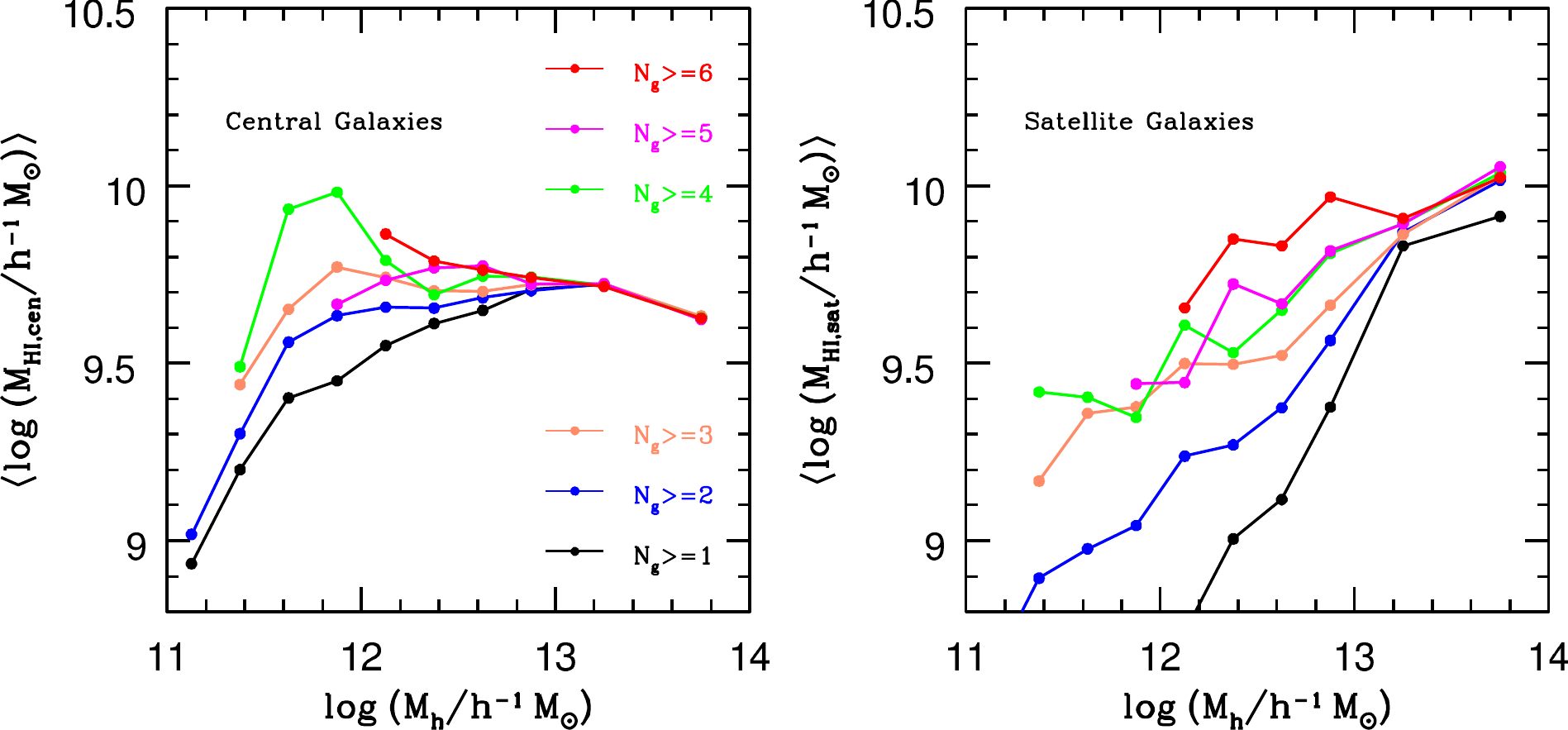}
	\caption{Similar to Figure~\ref{fig:hihm}, but for the contributions from the central galaxies (left panel) and from the summation of all satellite galaxies (right panel) in each halo. For clarity, the error bars of the measurements are omitted. }
	\label{fig:hihm_cs}
\end{figure*}
\begin{figure*}
	\centering
	\includegraphics[width=0.8\textwidth]{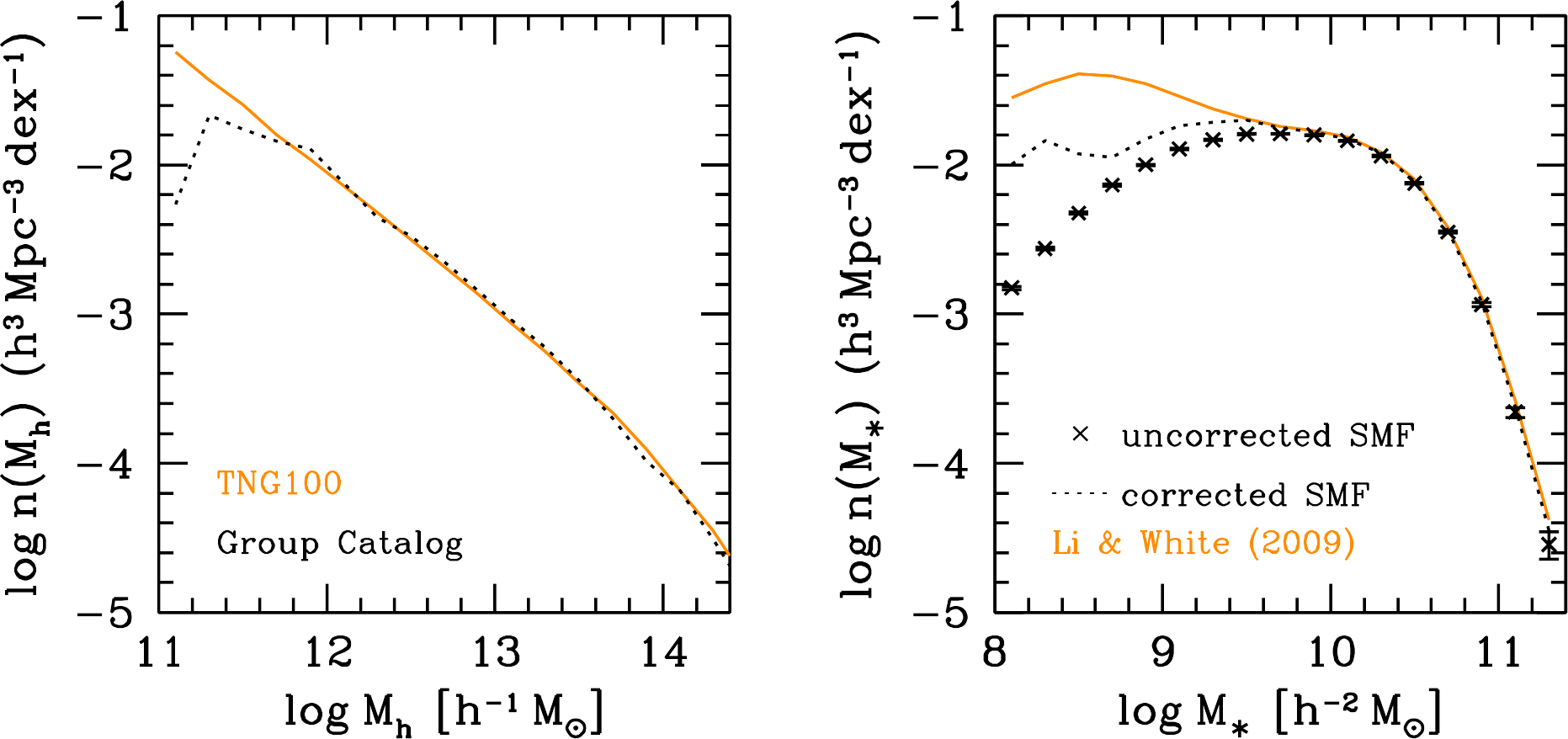}
	\caption{Left: comparison between the halo mass function $n(M_{\rm h})$ from the group catalog (dotted line) and that of the same cosmology directly obtained from the TNG100 simulation (solid line) of the IllustrisTNG simulation set. Right: observed galaxy stellar mass function in the group catalog (crosses), in comparison to the intrinsic measurement of \cite{Li+2009} (solid line). The stellar mass function corrected for the halo completeness is displayed as the dotted line.}
	\label{fig:halomf}
\end{figure*}
In order to separate the contribution to the total \hi\ mass into different components, we show in Figure~\ref{fig:hihm_cs} the average \hi\ mass from the central galaxies (left panel) and from the summation of all satellite galaxies (right panel) in halos of different masses, respectively. Measurements for halos of different richness thresholds are shown as different color lines. For clarity, the error bars of the measurements are omitted. As we have stacked the \hi\ spectra for the central galaxies, the contribution of all satellite galaxies in each halo mass bin to the total \hi\ mass is simply obtained by subtracting the \hi\ mass in the central galaxies from those of the halos. However, as we noted before,  stacking the \hi\ signals of the central galaxies would be contaminated by the confusion from the nearby satellite galaxies. Therefore, the measurements of the \hi\ mass in the centrals are upper limits, while those of the satellites are lower limits. We will discuss the effect of confusion in Section~\ref{subsubsec:confusion}.

The central galaxies dominate the contribution to the total \hi\ mass for low-mass halos, while the contribution from the satellite galaxies become comparable and even larger for halos of $M_{\rm h}>10^{12.5}\msunh$. We note that the bump feature in the \hi-halo mass relation is caused by the contribution from central galaxies, while the contribution from satellites monotonically increases with halo mass. At low halo masses, centrals with more surrounding satellites have higher \hi\ masses, but above $M_{\rm h}=10^{13}\msunh$ this trends disappears. Moreover, the effect of halo richness on $\langle\log M_{\rm HI}\rangle$ becomes increasingly smaller for larger $N_{\rm g}$.

The $\langle\log M_{\rm HI}\rangle$ for all satellite galaxies seems to follow power-law relations, with smaller slopes for higher halo richnesses. The trend with the halo richness is also similar to that of the central galaxies, but without the bump feature. In massive halos of $M_{\rm h}>10^{13}\msunh$, the majority of the \hi\ mass is contributed by the satellite galaxies. The values of $\langle M_{\rm HI}|M_{\rm h}\rangle$ for all halos (i.e., $N_{\rm g}\ge1$), as well as the corresponding values for the central galaxies, are displayed in Appendix~\ref{sec:app1}. 

\subsection{Systematic Effects}\label{subsec:robust}
Before discussing the implications of our measurements, we will first verify the results with several systematic tests. 

\subsubsection{Sample Completeness}\label{subsubsec:comp}
The first question is whether the trend of \hi-halo mass relation with the halo richness could be significantly affected by the incompleteness of the galaxy sample. Since the SDSS DR7 Main Galaxy Sample is flux-limited, we will miss faint galaxies at high redshifts. As we focus on the redshift range of $0.0025<z<0.06$ with available ALFALFA data, the sample is volume-limited for galaxies with a $r$-band absolute magnitude brighter than $M_{\rm r}\sim-18.8$ \citep[see e.g., Fig.~1 of][]{Guo+2015}, which corresponds to a galaxy stellar mass threshold around $10^{9.5}\msunhh$. Therefore, some of the gas-rich but optically faint galaxies in this redshift range might be missed in the group catalog due to the optical flux limit, which will affect the richness estimates of the halos. But the contribution of their \hi\ flux to the host halo are correctly included through the stacking method.

The galaxy sample completeness as a function of stellar mass, $C_{\rm g}(M_\ast)$, can be separated into the completeness of halos with a given mass $M_{\rm h}$, $C_{\rm h}(M_{\rm h})$, and the completeness of galaxies with a stellar mass $M_\ast$ in these halos, $C_{\rm g}(M_\ast|M_{\rm h})$. We are then dividing the missing galaxies into those in the missing halos and those in the observed halos. As there is no obvious selection bias of the halo population at a given mass for the flux-limited SDSS sample, the detected halos would be representative of the whole population. Then the halo richness estimate is only affected by $C_{\rm g}(M_\ast|M_{\rm h})$. The advantage of the group catalog is that we can estimate $C_{\rm g}(M_\ast)$ and $C_{\rm h}(M_{\rm h})$ directly from comparing to the intrinsic values from simulations and observations.

We show in the left panel of Figure~\ref{fig:halomf} the comparison between the halo mass function $n(M_{\rm h})$ from the group catalog (dotted line) and that of the same cosmology directly obtained from the TNG100 simulation (solid line) of the IllustrisTNG simulation set \citep{Marinacci+2018,Naiman+2018,Nelson+2018,Pillepich+2018,Springel+2018}. The dark matter halos in the group catalog is basically complete for $M_{\rm h}>10^{11.5}\msunh$, while the completeness decreases to around $1\%$ for halos of $M_{\rm h}\sim10^{11}\msunh$. The right panel of Figure~\ref{fig:halomf} shows the observed galaxy stellar mass function (SMF) in the group catalog (crosses), in comparison to the intrinsic measurement of \cite{Li+2009} (solid line) obtained by up-weighting galaxies with the maximum detectable volumes in $0.0025<z<0.06$. The observed galaxies are complete for $M_{\ast}>10^{10}\msunhh$, and the completeness $C_{\rm g}(M_\ast)$ decreases to 6\% for $M_{\ast}\sim10^8\msunhh$. 

As we have estimated the halo completeness, we can correct the effect of missing halos through weighting each galaxy by the corresponding value of $1/C_{\rm h}(M_{\rm h})$. The resulting SMF is shown as the dotted line. With the correction, we find that the galaxies in the observed halos are complete for $M_{\ast}>10^{9.5}\msunhh$. The stellar mass completeness is 60\% for $M_{\ast}\sim10^9\msunhh$ and decreases to 25\% for $M_{\ast}\sim10^8\msunhh$. Therefore, we are still missing some low-mass galaxies for the observed halos. As the low-mass halos are not likely to host many satellite galaxies, the majority of the missing galaxies would possibly be dwarf satellite galaxies in massive halos.

The value of a halo richness is then only meaningful with a stellar mass threshold, as we will always be missing those very low-mass galaxies. If we set the galaxy stellar mass threshold to be $M_{\ast}>10^7\msunhh$, more than $99.5\%$ of the observed halos would have the same richness values as provided in the group catalog. Thus, for fair comparisons with the theoretical models in the following sections, the halo richness $N_{\rm g}$ is defined as the number of galaxies with $M_{\ast}>10^7\msunhh$ in each halo.

\subsubsection{Confusion Correction}\label{subsubsec:confusion}
As the \hi\ spectral stacking technique simply cuts out a square box centered on the target position, it is inevitable that this includes some amount of confused \hi\ emissions from nearby objects in both the angular and radial directions. In order to estimate the effect of confusion, we can apply corrections to the \hi\ mass measurements for halos and central galaxies separately. 

\begin{figure*}
	\centering
	\includegraphics[width=0.8\textwidth]{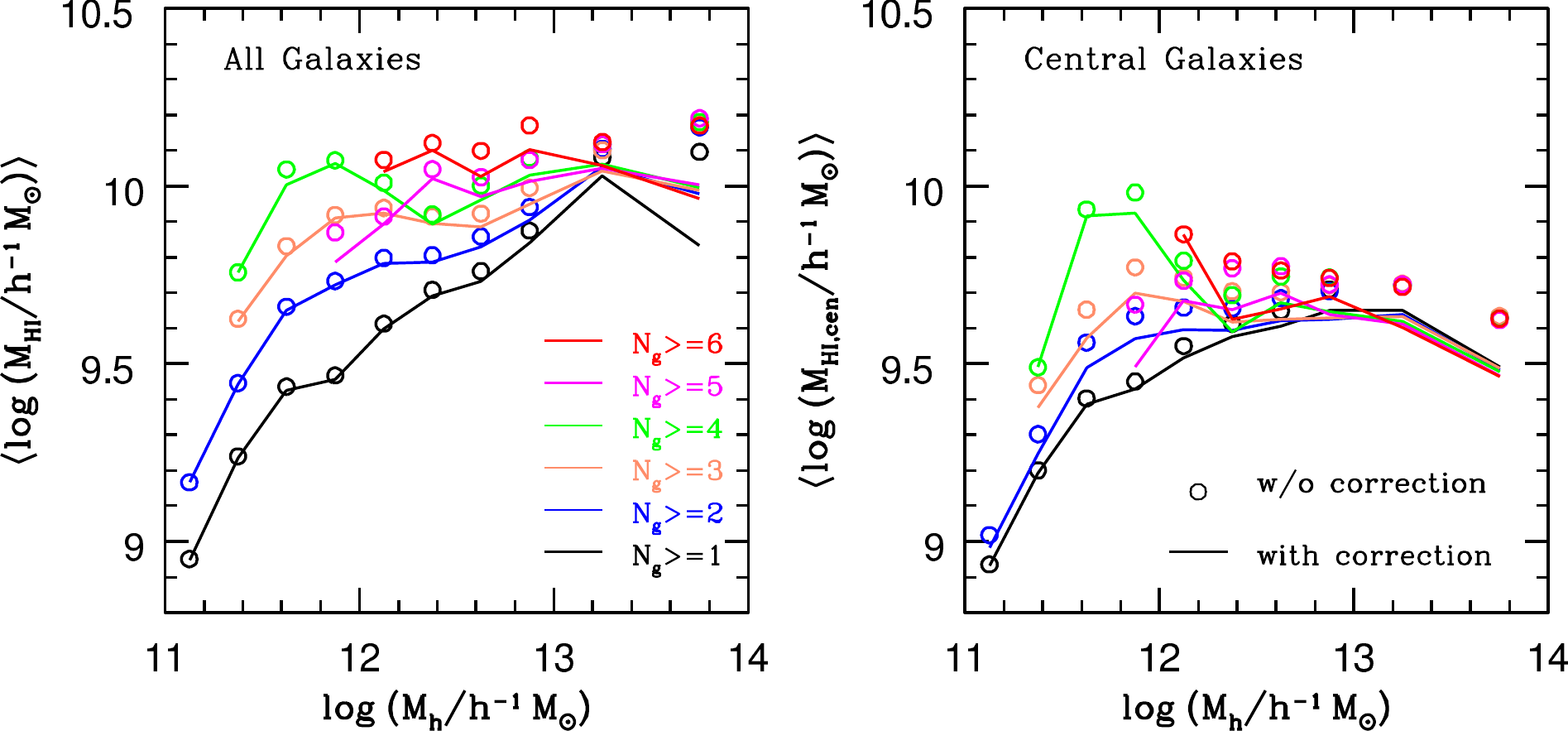}
	\caption{Effect of applying the confusion correction to the \hi-halo mass relations for the halos (left panel) and central galaxies (right panel). The open circles are the original measurements, while the solid lines are those with confusion corrections (see text for details). }
	\label{fig:confusion}
\end{figure*}
As noted in Section~\S\ref{sec:method}, the angular aperture size of the stacking for each halo is $\max(2r_{200}/D_{\rm A},8\arcmin)$, where $r_{200}$ and $D_{\rm A}$ are the virial radius and the angular diameter distance of the halo, respectively. In the radial direction, we integrate the \hi\ flux within $3\sigma$ of the peak of the stacked spectra for each halo mass bin. As shown in the stacked spectra in Appendix~\ref{sec:app2}, the $3\sigma$ velocity range of the low-mass halos of $M_{\rm h}\sim10^{11}\msunh$ is only around 150\,\kms, but increases to around 1000\,\kms for halos of $M_{\rm h}\sim10^{14}\msunh$. 

In order to estimate the contribution from confused \hi\ emission for the group halos, we employ a correction as follows. For each halo mass bin, we identify all the non-target halos from the group catalog which are within the apertures (angular and radial) used to stack the targets. Using the uncorrected relations from Figure~\ref{fig:hihm}, we can derive the average \hi\ mass in given halo mass and richness bins, $\langle M_{\rm HI}|M_{\rm h}, N_{\rm g}\rangle$, by subtracting the measurements between the two neighbouring richness threshold samples. We then estimate the \hi\ masses of these companion halos and thus how much they contribute to the stacked spectra in the relevant halo mass bin.

As halos in the group catalog are quite complete for $M_{\rm h}>10^{11.3}\msunh$, we can get a reliable estimate of the number of companion halos in each mass bin. We find that for $M_{\rm h}<10^{12.5}\msunh$, the average number of companion halos for each halo is typically smaller than $0.05$. It slightly increases to $0.3$ for $M_{\rm h}\sim10^{13}\msunh$ and becomes around $1.5$ for for $M_{\rm h}\sim10^{14}\msunh$. The correction to the \hi-halo mass relation is shown in the left panel of Figure~\ref{fig:confusion} for halos with different richnesses, with the circles for the original measurements and lines for the corrected ones. 

We note that our correction is an over-estimate of the real effect, as we simply assume the overlapping fraction between the confused halo and companions to be unity, different from the implementation of \cite{Fabello+2011}. Despite this, we find that the correction is basically minor for halos of $M_{\rm h}<10^{13}\msunh$, but becomes very significant for the largest halo mass bin of $10^{13.5}<M_{\rm h}<10^{14}\msunh$. As shown in Figure~\ref{fig:group_stacks}, due to the small number of available halos, the large noises in the stacked spectra of the most massive halos make the confusion correction less reliable. 

\begin{figure*}
	\centering
	\includegraphics[width=1.0\textwidth]{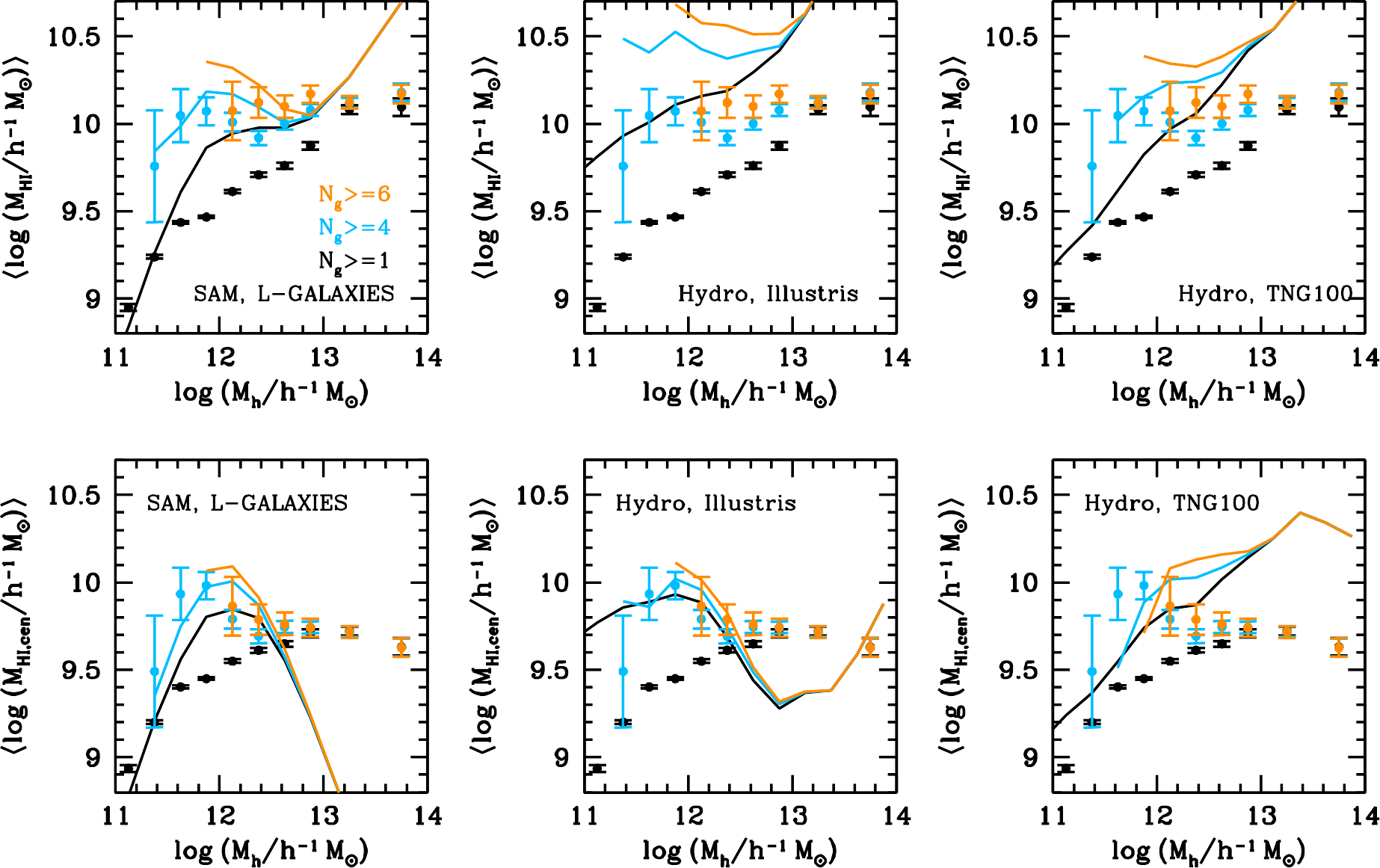}
	\caption{Comparisons between the \hi-halo mass relation measurements of observation (points) and different theoretical models (lines). The top and bottom panels are for the measurements of halo and central galaxies, respectively. The left, middle, and right panels are for the L-GALAXIES semi-analytical model, the Illustris, and IllustrisTNG hydrodynamical simulation models, respectively. For illustration, we only show three typical cases with the halo richnesses of $N_{\rm g}\ge1$ (black), $N_{\rm g}\ge4$ (blue), and $N_{\rm g}\ge6$ (orange).}
	\label{fig:hihm_model}
\end{figure*}
The confusion correction for the stacking of central galaxies is treated differently. The angular aperture size for central galaxies is $\max(200\,\rm{kpc}/D_{\rm A}, 8\arcmin)$, and we summed all emission within 300\,\kms of the peak values. Due to the large aperture size, the centrals are very likely to be confused with nearby satellite galaxies. The confusion effect would then become much larger for more massive halos with multiple satellites and this is compounded by the fact that satellites dominate the \hi\ content of massive halos. The average number of companion galaxies for each central is around $0.2$ for $M_{\rm h}\sim10^{12}\msunh$ and increases to $0.8$ for $M_{\rm h}\sim10^{13}\msunh$. So the confusion effect is much larger for central galaxies, compared to that of the halo. Moreover, the confusion effect of central galaxies would significantly increase with the halo richness, as expected. 

The confusion correction for the central galaxies is, however, hard to estimate. Because only less than 30\% of the individual galaxies in the group catalog have available \hi\ masses in the ALFALFA survey. As a lower limit of the confusion effect, we apply a minimal correction to the total \hi\ masses of the central galaxies by only subtracting the companions with measured \hi\ masses. The result is shown in the right panel of Figure~\ref{fig:confusion}. The correction is in general around 0.1~dex for halos of $M_{\rm h}>10^{12}\msunh$, but becomes much smaller for low-mass halos with a small richness. But the overall trend of $\langle M_{\rm HI}\rangle$ with the halo mass and richness is still quite similar to ones without correction. 

We can also estimate the \hi\ masses for those galaxies without ALFALFA detection by applying the scaling relations of the gas fraction with the optical galaxy properties \citep{Fabello+2012}, e.g., color and surface brightness, as in \cite{Zhang+2009} and \cite{Li+2012}. While such \hi\ mass estimators have large scatters of around 0.3~dex, we find that the confusion correction would reach to about 0.2~dex for halos of $M_{\rm h}\sim10^{12}\msunh$ and 0.3~dex for the most massive halos. It is thus essential to have reliable \hi\ mass estimates for the companions. Therefore, the stacking of the central galaxies can be deemed as upper limits of the total \hi\ mass content.  

We note that the halo mass estimate and central/satellite assignment in the group catalog are never perfect \citep{Campbell+2015}. There are inevitable measurement errors for the halo mass estimates. However, the typical halo mass error is around 0.2~dex, which corresponds to a small error of 0.067~dex for the halo virial radius, as $r_{200}\propto M_{\rm h}^{1/3}$. Moreover, the errors in the halo mass estimates are relatively compensated by our stacking of the halos.  While our measurements of the \hi-halo mass relation would potentially be slightly smoothed by the halo mass errors, the overall trend with the halo richness would not be affected. The mis-assignment of central and satellite galaxies is not a significant effect in our measurements, as we do not focus on the individual central and satellite galaxies. 

\subsection{Comparison to theoretical models}\label{subsec:model} 
While our direct measurements of the \hi-halo mass relations show a strong dependence on the halo richness, it is important to compare with the theoretical model predictions. We show in Figure~\ref{fig:hihm_model} the comparisons of $\langle M_{\rm HI}|M_{\rm h}\rangle$ for halos (top panels) and central galaxies (bottom panels) of different theoretical models. Our measurements are displayed as the symbols with error bars, while the model predictions are represented by the solid lines. We consider the L-GALAXIES semi-analytical model of \cite{Fu+2013} (left panels), the hydrodynamical simulations of Illustris \citep{Vogelsberger+2014} (middle panels) and IllustrisTNG (right panel). For illustration, we only show three typical cases with the halo richnesses of $N_{\rm g}\ge1$, $N_{\rm g}\ge4$, and $N_{\rm g}\ge6$, as symbols and lines of different colors.

The division of neutral hydrogen into \hi\ and H$_2$ in the Illustris and IllustrisTNG simulation models is implemented following \cite{Diemer+2018}, based on the \hi/H$_2$ transition model of \cite{Krumholz+2013}. It has been shown that while the Illustris model significantly over-predicts the abundance of \hi\ gas \citep{Guo+2017}, the IllustrisTNG model agrees much better with the observation \citep{Diemer+2019,Stevens+2019}. The \hi/H$_2$ transition in the L-GALAXIES model adopts the prescription of \cite{Blitz+2006}, which is based on the pressure in the local ISM.

\begin{figure*}
	\centering
	\includegraphics[width=0.8\textwidth]{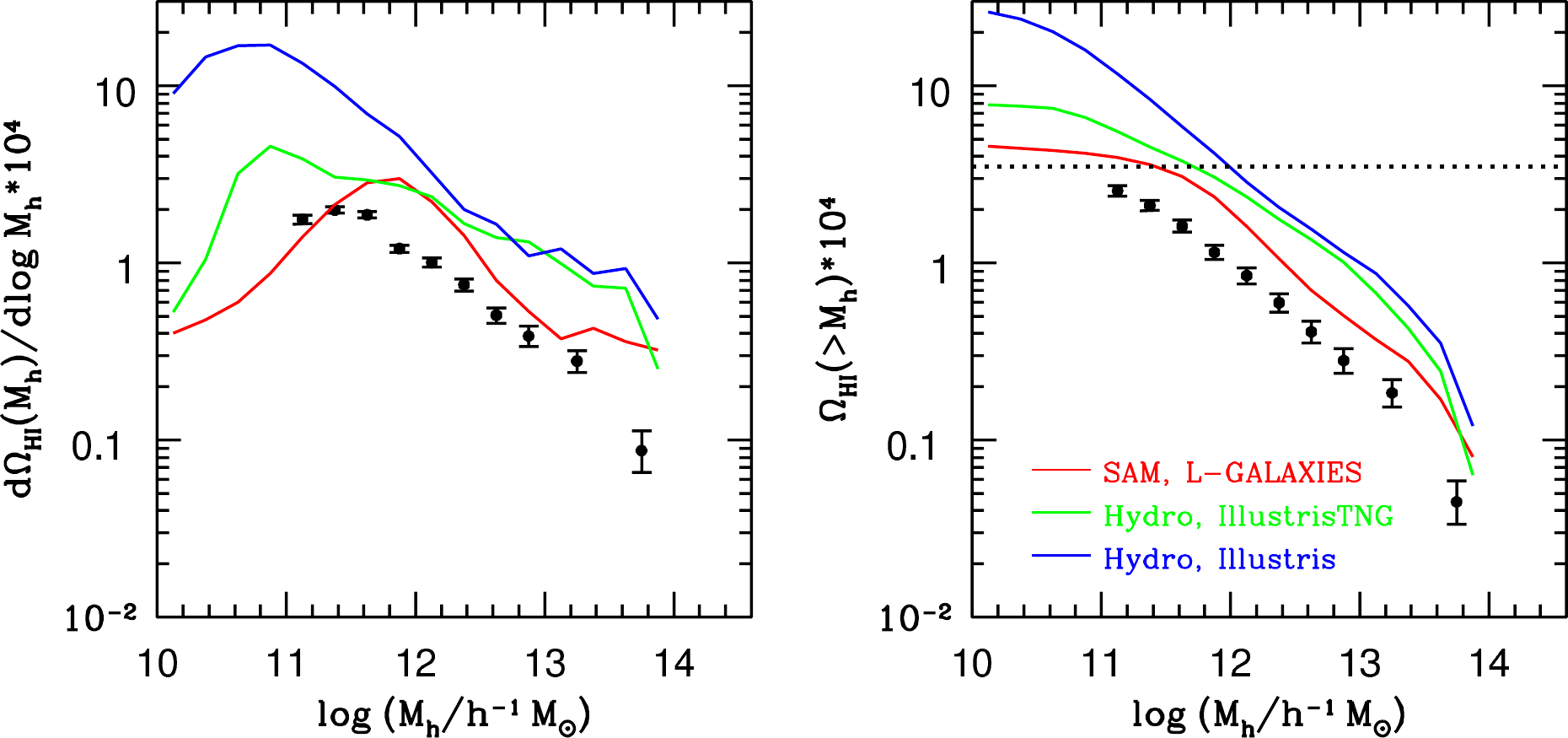}
	\caption{Fractional contribution to $\Omega_{\rm HI}$ of different halo mass bins $d\Omega_{\rm HI}/d\log M_{\rm h}$ (left panel), and the cumulative contribution of $\Omega_{\rm HI}(>M_{\rm h})$ above a given halo mass threshold (right panel). Our measurements are displayed as the points, while solid lines of different colors represent the three theoretical models. The dotted line in the right panel is the measurement from the \hi\ mass function of ALFALFA 100\% sample \citep{Jones+2018}.}
	\label{fig:omegahi}
\end{figure*}
Our stacking is based on the optical galaxies in the group catalog, but the different theoretical models have quite different galaxy stellar mass function predictions, which are not always consistent with the observation. Therefore, fair comparisons between observations and models should be made for galaxy samples above different stellar mass thresholds, but with the same number density. As stated before, our halo richness definition is consistent with a stellar mass threshold of $M_\ast>10^7\msunhh$. We calculate the ``complete'' sample number density by summing all galaxies with $M_\ast>10^7\msunhh$ from the SMF measurement of \cite{Li+2009}, as in the right panel of Figure~\ref{fig:halomf}. The resulting sample number density is $0.0696\,h^{-3}\,\rm{Mpc}^3$. The corresponding stellar mass thresholds for the L-GALAXIES, Illustris, and IllustrisTNG models are $10^{8.62}\msunhh$, $10^{8.75}\msunhh$, and $10^{8.30}\msunhh$, respectively. The stellar mass thresholds in different models are comparable, but much higher than the observation.

As shown in Figure~\ref{fig:hihm_model}, the same dependence on the halo richness is found in all the models, which confirms our finding. However, none of the models can well reproduce the observed \hi-halo mass relations. From the top panels of Figure~\ref{fig:hihm_model}, all three models significantly over-predict the total \hi\ mass for halos of $M_{\rm h}>10^{11.5}\msunh$ and $N_{\rm g}\ge1$. As we will show in the following, the excess of the \hi\ gas will result in the over-prediction of cosmic \hi\ gas density, $\Omega_{\rm HI}$.

As noted before, the central and satellite galaxies dominate over the contribution to the total \hi\ gas for halos below and above the transition mass $M_{\rm h}\sim10^{12.5}\msunh$, respectively. By comparing the top and bottom panels of Figure~\ref{fig:hihm_model}, we find that the L-GALAXIES model tends to agree with our measurements at the low halo mass end for both $N_{\rm g}\ge1$ and $N_{\rm g}\ge4$. However, the amount of \hi\ gas in the satellite galaxies in the massive halos is over-predicted. The situation is quite similar for the Illustris model, but the satellite contribution of the cold \hi\ gas in low-mass halos is too high. However, the improved IllustrisTNG hydrodynamical simulation model tends to put too little cool gas in the satellite galaxies in massive halos. The central galaxies contribute the majority of the \hi\ gas in halos of all masses. 

It is noteworthy that the bump feature of the \hi-halo mass relation is observed in all three models, but with different strengths. As explained in \cite{Baugh+2019}, the turn-over of $\langle M_{\rm HI}\rangle$ around halos of $M_{\rm h}~\sim10^{12}\msunh$ is generally assumed to be caused by the suppression of gas cooling by Active Galactic Nuclei (AGN) feedback. It seems that the low level of AGN feedback in the IllustrisTNG models makes the gas cooling too efficient in the massive halos. Therefore, our measurements can potentially be used to constrain the strength of AGN feedback. 

\subsection{Cosmic Neutral Hydrogen Gas Density}
With the \hi-halo mass relation, we can estimate the cosmic \hi\ gas density as, 
\begin{equation}
\Omega_{\rm HI}=\frac{1}{\rho_{\rm c}}\int \langle M_{\rm HI}|M_{\rm h}\rangle n(M_{\rm h})dM_{\rm h},
\end{equation}
where $n(M_{\rm h})$ is the intrinsic halo mass function and $\rho_{\rm c}$ is the critical density. The measurement of $\Omega_{\rm HI}$ can be directly compared with the one obtained from integrating the \hi\ mass function \citep[see e.g.,][]{Martin+2010,Jones+2018}. However, as we only measure $\langle M_{\rm HI}|M_{\rm h}\rangle$ for halos more massive than $10^{11}\msunh$, a proper extrapolation is necessary to obtain an accurate estimate of $\Omega_{\rm HI}$. The extrapolation for the \hi\ mass function is done using the Schechter function \citep{Martin+2010}. We reserve the investigation of the proper functional form for the \hi-halo mass relation to our future work.   

Using our current measurements, we can calculate the fractional contribution to $\Omega_{\rm HI}$ of different halo mass bins, $d\Omega_{\rm HI}/d\log M_{\rm h}$, and the cumulative contribution of $\Omega_{\rm HI}(>M_{\rm h})$ above a given halo mass threshold. We show (in Figure~\ref{fig:omegahi}) our measurements as the points, while solid lines of different colors represent the three theoretical models. The dotted line is the measurement from the \hi\ mass function of ALFALFA 100\% sample \citep{Jones+2018}. The predicted $\Omega_{\rm HI}$ is $3.5\times10^{-4}$ before applying the correction for \hi\ self-absorption, which is estimated to be 11\% in \cite{Jones+2018}. However, the self-absorption correction for the stacking of halos is very difficult to estimate, so we ignore the self-absorption correction in our measurements. 

We find that the cumulative \hi\ density $\Omega_{\rm HI}(M_{\rm h}>10^{11}\msunh)$ is $2.5\times10^{-4}$. Thus, there is still about 30\% of the total \hi\ gas in halos of $M_{\rm h}<10^{11}\msunh$, where the dwarf central galaxies would dominate the \hi\ mass contribution. Observation of these faint galaxies in future surveys would provide better constraints on $\Omega_{\rm HI}$. We note that all three theoretical models over-predict the cumulative contribution to $\Omega_{\rm HI}$, due to the overestimate of the \hi\ mass in massive halos. For the fractional contribution of $d\Omega_{\rm HI}/d\log M_{\rm h}$, the L-GALAXIES model shows better agreement than the two simulation models. The majority of the \hi\ mass is contributed by halos of $M_{\rm h}<10^{12}\msunh$, which is consistent with the predictions of \cite{Guo+2017} that most of the \hi-rich galaxies live in halos of $M_{\rm h}<10^{12}\msunh$.  

\section{Discussion}\label{sec:discussion}
The dependence of the total \hi\ mass on the halo richness in addition to the dependence on the halo mass reflects the halo assembly bias effect of \hi-rich galaxies. The halo assembly bias is generally referred to as the additional dependence of halo bias on properties of the halo formation history \citep[see e.g.,][]{Gao+2005,Gao+2007,Jing+2007}. As found in \cite{Guo+2017}, for halos of a given mass, those with a later formation time tend to host galaxies with a larger \hi\ mass. As shown in Figure~5 of \cite{Wechsler+2006}, halos with later formation times have higher richness values. Therefore, our finding of the richness dependence is fully consistent with the conclusion of \cite{Guo+2017}. They both confirm that the \hi\ mass is directly connected to the halo formation history. 

The commonly used indicators to characterize the halo formation history include the halo formation time, the spin parameter and the concentration parameter. The behavior of the halo assembly bias effect with the halo mass is quite different for the different indicators. For massive halos of $M_{\rm h}>10^{13}\msunh$, there is still a strong assembly bias effect with the halo concentration and spin parameter, while the effect with the halo formation time becomes much weaker \citep[e.g.,][]{Xu+2018,Sato-Polito+2019}, which is quite similar to the dependence of \hi\ mass on the halo richness. The similar behavior of \hi\ mass and the halo bias tends to indicate that the \hi\ mass is sensitive to the large-scale environment of the host halo. If true, we would expect to find the strong dependence of the \hi\ mass on the halo spin parameter for these massive halos, which could potentially be verified with the measurements of the \hi\ rotation curve.

However, although different halo properties are correlated with each other, the dependence of the total \hi\ gas mas on the different halo properties could potentially correspond to very different physical formation scenarios. It has previously been proposed that the \hi-rich galaxies tend to live in high-spin halos \citep[e.g.,][]{Huang+2012,Maddox+2015,Obreschkow+2016}. For example, \cite{Lutz+2018} used a sample of extremely \hi-rich galaxies and found a positive correlation between the gas ratio and the halo spin parameter. Galaxies in their sample have stellar masses in the range of $10^{10}\msun$--$10^{11}\msun$, which corresponds to a halo mass range around $10^{11.5}\msunh$--$10^{12.5}\msunh$ \citep{Moster+2010,Yang+2012}. If the \hi\ mass is indeed the indicator of the halo bias, the \hi\ mass would potentially have a much stronger dependence on the halo formation time than spin parameter in this mass range \citep[see e.g., Fig.~4 of][]{Sato-Polito+2019}. As shown in Figure~10 of \cite{Guo+2017}, the spin parameter is insufficient to explain the spatial clustering of the \hi-rich galaxies in the halo mass range of $10^{10}\msunh$--$10^{12}\msunh$.

The physical formation scenario of the \hi-rich galaxies is still under investigation. The \hi-rich galaxies can be formed from the recent gas accretion that increases the \hi\ reservoir, which is related to the halo formation time dependence. On the other hand, they can accrete their gas at an early time but the consumption of the cold \hi\ gas can be inefficient due to the high halo angular momentum that prevents the gas from collapsing and forming stars \citep{Obreschkow+2016,Lagos+2018,Stevens+2019}. While both factors can be important in the real formation scenario, they can play different roles at different stages. From combining the dependence of \hi\ mass on the different halo properties, the plausible scenario is that in group halos of $M_{\rm h}<10^{13}\msunh$, the recent gas accretion from the local halo environment is the dominant factor of the high \hi\ mass. This is also supported by the significant increase of total \hi\ mass in the satellite galaxies with the halo richness, as shown in Figure~\ref{fig:hihm_cs}. As the spin parameters of these late-forming halos are also relatively higher, the transition from atomic to molecular hydrogen may be further hindered by the high angular momentum and the lack of enough disk pressure \citep{Blitz+2006,Popping+2015,Lagos+2017}. 

As shown in Figure~\ref{fig:hihm_cs}, the growth of the total \hi\ mass is very efficient for halos of $M_{\rm h}<10^{11.8}\msunh$, where the contribution from ``cold mode'' of the gas accretion is significant. As shown in Figure~6 of \cite{Keres+2005}, the cold gas accretion becomes negligible in halos of $M_{\rm h}>10^{12}\msun$, consistent with our finding here. For more massive halos of $10^{11.8}\msunh<M_{\rm h}<10^{13}\msunh$, where the ``hot mode'' dominates, the supply of cold gas to the central galaxies is inefficient due to the virial shock-heating of the infalling gas \citep{Birnboim+2003,Keres+2005,Dekel+2006}. Therefore, the dependence of central \hi\ mass on the halo richness becomes weaker in this halo mass range. We note that the exact transition halo mass scale between the cold and hot mode dominance would depend on the definition of cold flows and the simulation models, typically varying from $10^{11.4}\msun$ to $10^{12}\msun$ \citep{Keres+2005,vandeVoort+2011,Correa+2018}.

However, as shown in Figure~13 of \cite{Keres+2005}, the transition mass scale is also dependent on the local galaxy number density. At $z=0$, the cold mode dominates for $n_{\rm gal}<0.2\,h^{3}\rm{Mpc}^{-3}$, which corresponds to a halo richness of $1.3$ for $M_{\rm h}\sim10^{12.5}\msunh$. Therefore, our results in Figures~\ref{fig:hihm} and~\ref{fig:hihm_cs} show that for halos with a richness smaller than 3, the cold mode accretion is important to the contribution of the \hi\ gas. As shown in the comparisons with the SAM of L-GALAXIES and the hydrodynamical simulations of Illustris and IllustrisTNG, where the virial shock-heating is included in the models, the effect of AGN feedback for halos of $M_{\rm h}>10^{12}\msunh$ is necessary to match the observed \hi\ mass. The strong dependence of the satellite \hi\ mass on the halo richness indicates that cold gas accretion in the outer parts of the halo is not significantly affected by the AGN feedback.  

For halos more massive than $10^{13}\msunh$, the hot accretion mode is not efficient \citep{Keres+2009} and the growth of halos is dominated by the mergers \citep[see e.g., Fig.~5 of][]{Genel+2010}. The halos with different formation times have similar large-scale environments, reflected by the insensitivity of halo bias on the formation time. The amount of accreted cold gas is then similar for these halos, manifested by the independence of \hi\ mass on the halo richness. As the effect of AGN feedback increases with the halo mass, the average cold gas mass of the central galaxy is then decreasing with the halo mass. The halo spin parameter would then potentially play a dominant role in determining the total \hi\ mass in these halos, as the consumption of \hi\ gas is lower for higher-spin halos.

\begin{figure}
	\centering
	\includegraphics[width=0.9\columnwidth]{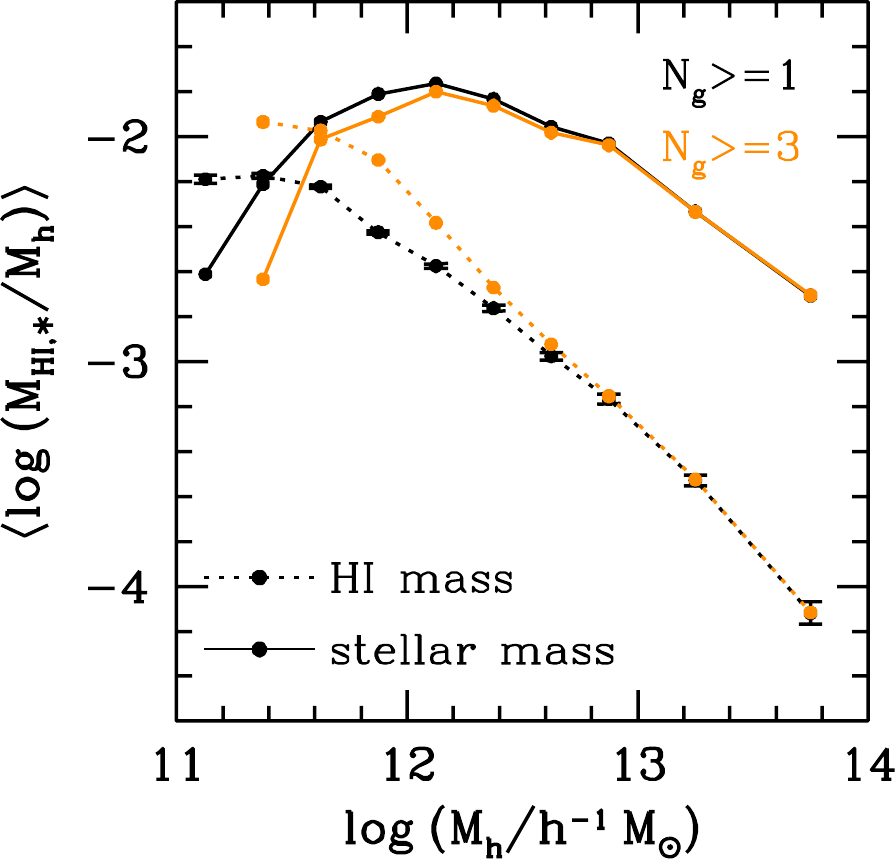}
	\caption{Comparisons between the gas fraction $M_{\rm HI}/M_{\rm h}$ (points with dotted lines) and the stellar mass fraction $M_\ast/M_{\rm h}$ (points with solid lines), for central galaxies in halos with $N_{\rm g}\ge1$ (black lines) and $N_{\rm g}\ge3$ (orange lines). }
	\label{fig:hihm_stellar}
\end{figure}
To study the connection between \hi\ gas and star formation, we show in Figure~\ref{fig:hihm_stellar} the comparisons between the gas fraction $M_{\rm HI}/M_{\rm h}$ (points with dotted lines) and the stellar mass fraction $M_\ast/M_{\rm h}$ (points with solid lines), for central galaxies in halos with $N_{\rm g}\ge1$ (black lines) and $N_{\rm g}\ge3$ (orange lines).  The peak of the gas fraction happens around $M_{\rm h}\sim10^{11.5}\msunh$, while that of the stellar mass fraction is at $M_{\rm h}\sim10^{12.1}\msunh$. It indicates that the AGN feedback starts to be effective from a halo mass of $10^{11.5}\msunh$. While the gas fraction is decreasing with the halo mass from $10^{11.5}\msunh$ to $10^{12}\msunh$, the total \hi\ mass is still rapidly increasing through the smooth accretion up to $M_{\rm h}=10^{11.8}\msunh$, where the virial shock-heating starts to heat the infalling gas \citep{Dekel+2006}. The accretion of \hi\ gas in this halo mass range contributes to the increase of galaxy stellar mass before reaching the peak at $M_{\rm h}\sim10^{12.1}\msunh$. We note that the stellar mass fraction of halos with $N_{\rm g}\ge3$ is slightly smaller than those with $N_{\rm g}\ge1$. The late accretion of cold gas in the high-richness halos may cause less \hi\ gas to be converted into stars.

In summary, the formation scenario of the \hi-rich galaxies involves complicated physical processes. Both the halo mass and the halo formation history are important in the various processes. Further investigations in the semi-analytical models and the hydrodynamical simulations to reproduce the observed \hi-halo mass relation would help understand their formation and evolution.

\section{Conclusions}\label{sec:summary}
In this paper, we have accurately measured the total \hi\ mass in halos of different masses by stacking \hi\ spectra of entire groups within the ALFALFA survey. Using the galaxy group catalog constructed from the optical survey of SDSS DR7, we are able to reliably determine both the halo mass and the halo membership, thereby constraining the \hi-halo mass relation for halos in a broad mass range from $10^{11}\msunh$ to $10^{14}\msunh$. It provides important constraints to the formation of the \hi-rich galaxies. 

Our main conclusions are summarized as follows.

\begin{itemize}
   \item Total \hi\ mass is not a single monotonically increasing function of halo mass. There is a bump in the function around $M_{\rm h}\sim10^{12}\msunh$. 
   
   \item The contribution to the total \hi\ mass is dominated by the central galaxies for halos of $M_{\rm h}<10^{12}\msunh$. Above this mass, satellites make the dominant contribution. 
   
   \item The \hi\ mass of a halo is not only a function of halo mass. There is a significant secondary dependence on richness, with richer halos having higher \hi\ masses. This secondary dependence is strongest for low-mass halos and completely absent for the most massive halos. 
   
   \item The bump in the \hi-halo mass relation is the result of a sharp dip in the \hi\ mass of centrals (in halos with $N_{\rm g}>=2$) at a halo mass of $M_{\rm h}\sim10^{12}\msunh$. The total \hi\ mass in satellite galaxies, on the other hand, monotonically increases with halo mass.
	
	\item We compare our measurements to the L-GALAXIES SAM and the hydrodynamical simulation models of Illustris and IllustrisTNG. We find that all of the models over-predict the abundance of \hi\ gas for halos of $M_{\rm h}>10^{11.5}\msunh$. The drop of \hi\ mass in central galaxies is observed in different models with various levels. The strength of AGN feedback in the theoretical models is the key part to reproduce the observation. 
	
	\item Our accurate measurements of the \hi-halo mass relation implies that the formation of the \hi\ gas in the halo can be divided into three phases. The smooth cold gas accretion is driving the growth of \hi\ mass in halos of $M_{\rm h}<10^{11.8}\msunh$, with late-forming halos having more cold \hi\ gas accreted. The virial halo shock-heating and AGN feedback will reduce the cold gas supply in halos of $10^{11.8}\msunh<M_{\rm h}<10^{13}\msunh$. The \hi\ mass in halos more massive than $10^{13}\msunh$ generally grows by mergers, where the dependence on halo richness and formation time becomes much weaker.
	
\end{itemize}

\acknowledgments
We thank the anonymous reviewer for the helpful comments that improve the presentation of this paper. 
We acknowledge the work of the entire ALFALFA team for observing, flagging and performing signal extraction. We thank Benedikt Diemer for providing the gas properties from the Illustris and IllustrisTNG simulations. We also thank Carlton Baugh, Cheng Li, Zheng Zheng, Claudia Lagos, Garima Chauhan and Adam Stevens for helpful discussions.

This work is supported by the National Key R\&D Program of China (grant No. 2018YFA0404503), national science foundation of China (Nos. 11773049, 11833005, 11828302, 11922305), and Spanish Science Ministry ``Centro de Excelencia Severo Ochoa'' program under grant SEV-2017-0709. MGJ is supported by a Juan de la Cierva formaci\'{o}n fellowship (FJCI-2016-29685). MGJ also acknowledges support from the grants AYA2015-65973-C3-1-R and RTI2018-096228-B-C31 (MINECO/FEDER, UE). MPH acknowledges support from NSF/AST-1714828 and the Brinson Foundation. JF acknowledges the support by the Youth innovation Promotion Association CAS and Shanghai Committee of Science and Technology grant (No.19ZR1466700). This research was supported by the Munich Institute for Astro- and Particle Physics (MIAPP) which is funded by the Deutsche Forschungsgemeinschaft (DFG, German Research Foundation) under Germany's Excellence Strategy ¨C EXC-2094 ¨C 390783311.

Funding for the SDSS and SDSS-II has been provided by the Alfred P. Sloan Foundation, the Participating Institutions, the National Science Foundation, the U.S. Department of Energy, the National Aeronautics and Space Administration, the Japanese Monbukagakusho, the Max Planck Society, and the Higher Education Funding Council for England. The SDSS Web Site is http://www.sdss.org/.

\begin{deluxetable*}{llcccccc}
	\centering	
	\tablecaption{Measurements of the \hi-halo mass relation\label{tab:hihm}} 
	\tablehead{$\log M_{\rm h}$ range & $N_{\rm halo}$ & $\langle\log M_{\rm HI}\rangle_{N_{\rm g}\ge1}$ & $\langle\log M_{\rm HI}\rangle_{N_{\rm g}\ge2}$ &
		$\langle\log M_{\rm HI}\rangle_{N_{\rm g}\ge3}$ & $\langle\log M_{\rm HI}\rangle_{N_{\rm g}\ge4}$ & $\langle\log M_{\rm HI}\rangle_{N_{\rm g}\ge5}$ & $\langle\log M_{\rm HI}\rangle_{N_{\rm g}\ge6}$ }
	\startdata
	$[11.0,11.25]$ & 949  & $ 8.951\pm0.020$ & $ 9.167\pm0.114$ &      \nodata     &      \nodata     & \nodata          & \nodata \\
	$[11.25,11.5]$ & 3494 & $ 9.239\pm0.011$ & $ 9.445\pm0.036$ & $ 9.626\pm0.106$ & $ 9.757\pm0.320$ & \nodata          & \nodata \\
	$[11.5,11.75]$ & 5472 & $ 9.436\pm0.008$ & $ 9.660\pm0.024$ & $ 9.831\pm0.070$ & $10.046\pm0.150$ & \nodata          & \nodata  \\
	$[11.75,12.0]$ & 7021 & $ 9.467\pm0.007$ & $ 9.733\pm0.016$ & $ 9.918\pm0.039$ & $10.072\pm0.081$ & $ 9.869\pm0.156$ & \nodata  \\
	$[12.0,12.25]$ & 3903 & $ 9.613\pm0.010$ & $ 9.798\pm0.016$ & $ 9.938\pm0.029$ & $10.009\pm0.053$ & $ 9.914\pm0.094$ & $10.074\pm0.167$ \\
	$[12.25,12.5]$ & 2074 & $ 9.708\pm0.014$ & $ 9.805\pm0.018$ & $ 9.914\pm0.027$ & $ 9.920\pm0.041$ & $10.047\pm0.059$ & $10.121\pm0.088$ \\
	$[12.5,12.75]$ & 1375 & $ 9.760\pm0.017$ & $ 9.858\pm0.021$ & $ 9.922\pm0.026$ & $10.001\pm0.035$ & $10.025\pm0.047$ & $10.099\pm0.064$ \\
	$[12.75,13.0]$ & 803  & $ 9.875\pm0.022$ & $ 9.941\pm0.025$ & $ 9.994\pm0.028$ & $10.078\pm0.034$ & $10.073\pm0.042$ & $10.170\pm0.050$ \\
	$[13.0,13.5]$  & 646  & $10.081\pm0.024$ & $10.104\pm0.026$ & $10.100\pm0.027$ & $10.117\pm0.029$ & $10.117\pm0.032$ & $10.124\pm0.035$ \\
	$[13.5,14.0]$  & 147  & $10.096\pm0.050$ & $10.165\pm0.051$ & $10.175\pm0.051$ & $10.180\pm0.051$ & $10.191\pm0.052$ & $10.170\pm0.053$ 
	\enddata   						
	
	\tablecomments{Measurements of $\langle M_{\rm HI}|M_{\rm h}\rangle$ for halos with different richness $N_{\rm g}$. All masses are in units of $\msunh$. The range of halo mass and total number of halos $N_{\rm halo}$ used in stacking of each mass bin for $N_{\rm g}\ge1$ are also displayed.}
\end{deluxetable*}

\begin{deluxetable*}{lcccccc}
	\centering	
	\tablecaption{Measurements of the \hi-halo mass relation for central galaxies\label{tab:hihm_cen}} 
	\tablehead{$\log M_{\rm h}$ range & $\langle\log M_{\rm HI,cen}\rangle_{N_{\rm g}\ge1}$ & $\langle\log M_{\rm HI,cen}\rangle_{N_{\rm g}\ge2}$ &
		$\langle\log M_{\rm HI,cen}\rangle_{N_{\rm g}\ge3}$ & $\langle\log M_{\rm HI,cen}\rangle_{N_{\rm g}\ge4}$ & $\langle\log M_{\rm HI,cen}\rangle_{N_{\rm g}\ge5}$ & $\langle\log M_{\rm HI,cen}\rangle_{N_{\rm g}\ge6}$ }
	\startdata
	$[11.0,11.25]$ & $ 8.936\pm0.020$ & $ 9.018\pm0.114$ &      \nodata     &      \nodata     &      \nodata     &      \nodata     \\
	$[11.25,11.5]$ & $ 9.200\pm0.010$ & $ 9.302\pm0.036$ & $ 9.440\pm0.106$ & $ 9.490\pm0.320$ &      \nodata     &      \nodata     \\
	$[11.5,11.75]$ & $ 9.402\pm0.008$ & $ 9.559\pm0.024$ & $ 9.652\pm0.070$ & $ 9.934\pm0.150$ &      \nodata     &      \nodata     \\
	$[11.75,12.0]$ & $ 9.450\pm0.007$ & $ 9.634\pm0.016$ & $ 9.771\pm0.038$ & $ 9.982\pm0.080$ & $ 9.666\pm0.155$ &      \nodata      \\
	$[12.0,12.25]$ & $ 9.550\pm0.010$ & $ 9.658\pm0.015$ & $ 9.742\pm0.029$ & $ 9.790\pm0.053$ & $ 9.734\pm0.094$ & $ 9.864\pm0.167$ \\
	$[12.25,12.5]$ & $ 9.612\pm0.013$ & $ 9.655\pm0.018$ & $ 9.704\pm0.027$ & $ 9.693\pm0.041$ & $ 9.769\pm0.058$ & $ 9.788\pm0.087$ \\
	$[12.5,12.75]$ & $ 9.649\pm0.017$ & $ 9.684\pm0.021$ & $ 9.702\pm0.026$ & $ 9.746\pm0.035$ & $ 9.774\pm0.047$ & $ 9.762\pm0.064$ \\
	$[12.75,13.0]$ & $ 9.709\pm0.022$ & $ 9.704\pm0.025$ & $ 9.721\pm0.028$ & $ 9.743\pm0.034$ & $ 9.723\pm0.042$ & $ 9.741\pm0.050$  \\
	$[13.0,13.5]$  & $ 9.722\pm0.024$ & $ 9.723\pm0.026$ & $ 9.725\pm0.027$ & $ 9.719\pm0.029$ & $ 9.723\pm0.032$ & $ 9.716\pm0.035$  \\
	$[13.5,14.0]$  & $ 9.632\pm0.050$ & $ 9.629\pm0.051$ & $ 9.634\pm0.052$ & $ 9.628\pm0.052$ & $ 9.623\pm0.053$ & $ 9.627\pm0.053$ 
	\enddata   						
	
	\tablecomments{Similar to Table~\ref{tab:hihm}, but for the central galaxies.}
\end{deluxetable*}
\appendix
\section{Measurements of \hi-halo mass relation}\label{sec:app1}
We list in Tables~\ref{tab:hihm} and~\ref{tab:hihm_cen} the measurements of the \hi-halo mass relation for the halos and central galaxies, respectively. The number of halos in each mass bin used in the stacking is also displayed. We note that in the final stacking we only include part of the halos in each bin due to the effects of low S/N, failed spectra or overlapping with the survey boundary.

\section{Stacked Spectra} \label{sec:app2}
\begin{figure*}
	\centering
	\includegraphics[width=\textwidth]{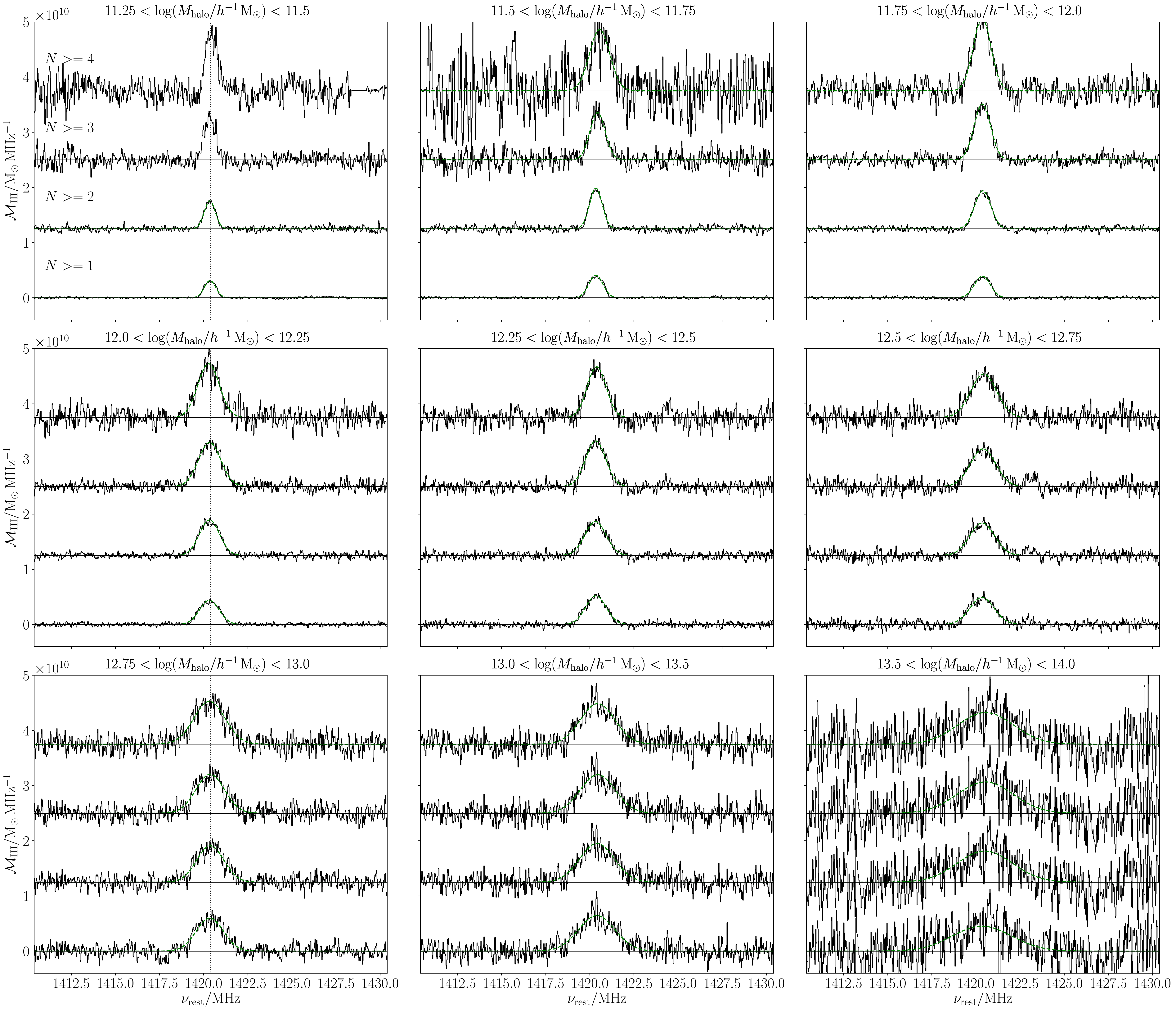}
	\caption{Stacks of groups divided into halo mass bins in separate panels. The vertical offset corresponds to the membership thresholds ranging from a minimum of one to four members (increasing upwards). The green dashed lines show the Gaussian fits to the profiles where appropriate.}
	\label{fig:group_stacks}
\end{figure*}

\begin{figure*}
	\centering
	\includegraphics[width=\textwidth]{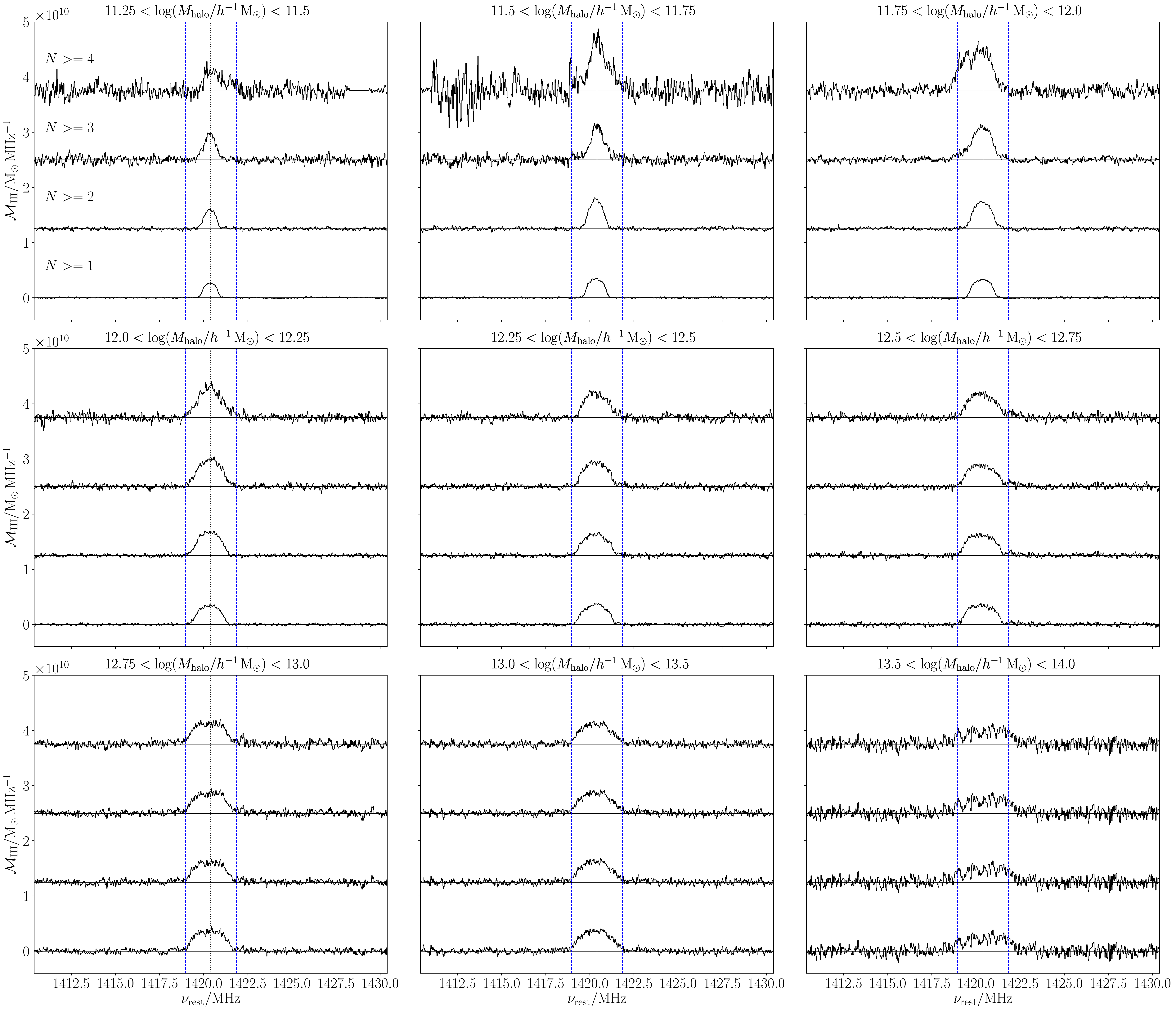}
	\caption{Stacks of group centrals divided into halo mass bins in separate panels. The vertical offset corresponds to the membership thresholds ranging from a minimum of 1 to 4 members (increasing upwards). The vertical blue dashed lines show the range the flux was integrated over to estimate the total \hi\ mass of the centrals in each stack.}
	\label{fig:central_stacks}
\end{figure*}

We show the stacked spectra for halos and central galaxies in Figures~\ref{fig:group_stacks} and~\ref{fig:central_stacks}. We only show the results for halos in the mass range of $10^{11.25}$--$10^{14}\msunh$, with different richness values from a minimum of one member to four members.

\bibliographystyle{aasjournal}
\bibliography{references}

\begin{thebibliography}{}
\expandafter\ifx\csname natexlab\endcsname\relax\def\natexlab#1{#1}\fi
\providecommand{\url}[1]{\href{#1}{#1}}
\providecommand{\dodoi}[1]{doi:~\href{http://doi.org/#1}{\nolinkurl{#1}}}
\providecommand{\doeprint}[1]{\href{http://ascl.net/#1}{\nolinkurl{http://ascl.net/#1}}}
\providecommand{\doarXiv}[1]{\href{https://arxiv.org/abs/#1}{\nolinkurl{https://arxiv.org/abs/#1}}}

\bibitem[{{Ai} \& {Zhu}(2018)}]{Ai+2018}
{Ai}, M., \& {Zhu}, M. 2018, \apj, 862, 48, \dodoi{10.3847/1538-4357/aac9b7}

\bibitem[{{Albareti} {et~al.}(2017){Albareti}, {Allende Prieto}, {Almeida}, \&
  et~al.}]{Albareti+2017}
{Albareti}, F.~D., {Allende Prieto}, C., {Almeida}, A., \& et~al. 2017, \apjs,
  233, 25, \dodoi{10.3847/1538-4365/aa8992}

\bibitem[{{Barnes} {et~al.}(2001){Barnes}, {Staveley-Smith}, {de Blok},
  {Oosterloo}, {Stewart}, {Wright}, {Banks}, {Bhathal}, {Boyce}, {Calabretta},
  {Disney}, {Drinkwater}, {Ekers}, {Freeman}, {Gibson}, {Green}, {Haynes}, {te
  Lintel Hekkert}, {Henning}, {Jerjen}, {Juraszek}, {Kesteven}, {Kilborn},
  {Knezek}, {Koribalski}, {Kraan-Korteweg}, {Malin}, {Marquarding}, {Minchin},
  {Mould}, {Price}, {Putman}, {Ryder}, {Sadler}, {Schr{\"o}der}, {Stootman},
  {Webster}, {Wilson}, \& {Ye}}]{Barnes+2001}
{Barnes}, D.~G., {Staveley-Smith}, L., {de Blok}, W.~J.~G., {et~al.} 2001,
  \mnras, 322, 486, \dodoi{10.1046/j.1365-8711.2001.04102.x}

\bibitem[{{Barnes} \& {Haehnelt}(2014)}]{Barnes+2014}
{Barnes}, L.~A., \& {Haehnelt}, M.~G. 2014, \mnras, 440, 2313,
  \dodoi{10.1093/mnras/stu445}

\bibitem[{{Baugh} {et~al.}(2019){Baugh}, {Gonzalez-Perez}, {Lagos}, {Lacey},
  {Helly}, {Jenkins}, {Frenk}, {Benson}, {Bower}, \& {Cole}}]{Baugh+2019}
{Baugh}, C.~M., {Gonzalez-Perez}, V., {Lagos}, C. d.~P., {et~al.} 2019, \mnras,
  483, 4922, \dodoi{10.1093/mnras/sty3427}

\bibitem[{{Birnboim} \& {Dekel}(2003)}]{Birnboim+2003}
{Birnboim}, Y., \& {Dekel}, A. 2003, \mnras, 345, 349,
  \dodoi{10.1046/j.1365-8711.2003.06955.x}

\bibitem[{{Blitz} \& {Rosolowsky}(2006)}]{Blitz+2006}
{Blitz}, L., \& {Rosolowsky}, E. 2006, \apj, 650, 933, \dodoi{10.1086/505417}

\bibitem[{{Brown} {et~al.}(2015){Brown}, {Catinella}, {Cortese}, {Kilborn},
  {Haynes}, \& {Giovanelli}}]{Brown+2015}
{Brown}, T., {Catinella}, B., {Cortese}, L., {et~al.} 2015, \mnras, 452, 2479,
  \dodoi{10.1093/mnras/stv1311}

\bibitem[{{Brown} {et~al.}(2017){Brown}, {Catinella}, {Cortese}, {Lagos},
  {Dav{\'e}}, {Kilborn}, {Haynes}, {Giovanelli}, \&
  {Rafieferantsoa}}]{Brown+2017}
---. 2017, \mnras, 466, 1275, \dodoi{10.1093/mnras/stw2991}

\bibitem[{{Campbell} {et~al.}(2015){Campbell}, {van den Bosch}, {Hearin},
  {Padmanabhan}, {Berlind}, {Mo}, {Tinker}, \& {Yang}}]{Campbell+2015}
{Campbell}, D., {van den Bosch}, F.~C., {Hearin}, A., {et~al.} 2015, \mnras,
  452, 444, \dodoi{10.1093/mnras/stv1091}

\bibitem[{{Catinella} {et~al.}(2010){Catinella}, {Schiminovich}, {Kauffmann},
  {Fabello}, {Wang}, {Hummels}, {Lemonias}, {Moran}, {Wu}, {Giovanelli},
  {Haynes}, {Heckman}, {Basu-Zych}, {Blanton}, {Brinchmann}, {Budav{\'a}ri},
  {Gon{\c{c}}alves}, {Johnson}, {Kennicutt}, {Madore}, {Martin}, {Rich},
  {Tacconi}, {Thilker}, {Wild}, \& {Wyder}}]{Catinella+2010}
{Catinella}, B., {Schiminovich}, D., {Kauffmann}, G., {et~al.} 2010, \mnras,
  403, 683, \dodoi{10.1111/j.1365-2966.2009.16180.x}

\bibitem[{{Correa} {et~al.}(2018){Correa}, {Schaye}, {Wyithe}, {Duffy},
  {Theuns}, {Crain}, \& {Bower}}]{Correa+2018}
{Correa}, C.~A., {Schaye}, J., {Wyithe}, J. S.~B., {et~al.} 2018, \mnras, 473,
  538, \dodoi{10.1093/mnras/stx2332}

\bibitem[{{Dekel} \& {Birnboim}(2006)}]{Dekel+2006}
{Dekel}, A., \& {Birnboim}, Y. 2006, \mnras, 368, 2,
  \dodoi{10.1111/j.1365-2966.2006.10145.x}

\bibitem[{{Delhaize} {et~al.}(2013){Delhaize}, {Meyer}, {Staveley-Smith}, \&
  {Boyle}}]{Delhaize+2013}
{Delhaize}, J., {Meyer}, M.~J., {Staveley-Smith}, L., \& {Boyle}, B.~J. 2013,
  \mnras, 433, 1398, \dodoi{10.1093/mnras/stt810}

\bibitem[{{Di Matteo} {et~al.}(2005){Di Matteo}, {Springel}, \&
  {Hernquist}}]{DiMatteo+2005}
{Di Matteo}, T., {Springel}, V., \& {Hernquist}, L. 2005, \nat, 433, 604,
  \dodoi{10.1038/nature03335}

\bibitem[{{Diemer} {et~al.}(2018){Diemer}, {Stevens}, {Forbes}, {Marinacci},
  {Hernquist}, {Lagos}, {Sternberg}, {Pillepich}, {Nelson}, {Popping},
  {Villaescusa-Navarro}, {Torrey}, \& {Vogelsberger}}]{Diemer+2018}
{Diemer}, B., {Stevens}, A. R.~H., {Forbes}, J.~C., {et~al.} 2018, \apjs, 238,
  33, \dodoi{10.3847/1538-4365/aae387}

\bibitem[{{Diemer} {et~al.}(2019){Diemer}, {Stevens}, {Lagos}, {Calette},
  {Tacchella}, {Hernquist}, {Marinacci}, {Nelson}, {Pillepich},
  {Rodriguez-Gomez}, {Villaescusa-Navarro}, \& {Vogelsberger}}]{Diemer+2019}
{Diemer}, B., {Stevens}, A. R.~H., {Lagos}, C. d.~P., {et~al.} 2019, \mnras,
  487, 1529, \dodoi{10.1093/mnras/stz1323}

\bibitem[{{Fabello} {et~al.}(2011){Fabello}, {Catinella}, {Giovanelli},
  {Kauffmann}, {Haynes}, {Heckman}, \& {Schiminovich}}]{Fabello+2011}
{Fabello}, S., {Catinella}, B., {Giovanelli}, R., {et~al.} 2011, \mnras, 411,
  993, \dodoi{10.1111/j.1365-2966.2010.17742.x}

\bibitem[{{Fabello} {et~al.}(2012){Fabello}, {Kauffmann}, {Catinella}, {Li},
  {Giovanelli}, \& {Haynes}}]{Fabello+2012}
{Fabello}, S., {Kauffmann}, G., {Catinella}, B., {et~al.} 2012, \mnras, 427,
  2841, \dodoi{10.1111/j.1365-2966.2012.22088.x}

\bibitem[{{Fu} {et~al.}(2013){Fu}, {Kauffmann}, {Huang}, {Yates}, {Moran},
  {Heckman}, {Dav{\'e}}, {Guo}, \& {Henriques}}]{Fu+2013}
{Fu}, J., {Kauffmann}, G., {Huang}, M.-l., {et~al.} 2013, \mnras, 434, 1531,
  \dodoi{10.1093/mnras/stt1117}

\bibitem[{{Gao} {et~al.}(2005){Gao}, {Springel}, \& {White}}]{Gao+2005}
{Gao}, L., {Springel}, V., \& {White}, S.~D.~M. 2005, \mnras, 363, L66,
  \dodoi{10.1111/j.1745-3933.2005.00084.x}

\bibitem[{{Gao} \& {White}(2007)}]{Gao+2007}
{Gao}, L., \& {White}, S.~D.~M. 2007, \mnras, 377, L5,
  \dodoi{10.1111/j.1745-3933.2007.00292.x}

\bibitem[{{Genel} {et~al.}(2010){Genel}, {Bouch{\'e}}, {Naab}, {Sternberg}, \&
  {Genzel}}]{Genel+2010}
{Genel}, S., {Bouch{\'e}}, N., {Naab}, T., {Sternberg}, A., \& {Genzel}, R.
  2010, \apj, 719, 229, \dodoi{10.1088/0004-637X/719/1/229}

\bibitem[{{Ger{\'e}b} {et~al.}(2015){Ger{\'e}b}, {Morganti}, {Oosterloo},
  {Hoppmann}, \& {Staveley-Smith}}]{Gereb+2015}
{Ger{\'e}b}, K., {Morganti}, R., {Oosterloo}, T.~A., {Hoppmann}, L., \&
  {Staveley-Smith}, L. 2015, \aap, 580, A43,
  \dodoi{10.1051/0004-6361/201424810}

\bibitem[{{Giovanelli} {et~al.}(2005){Giovanelli}, {Haynes}, {Kent},
  {Perillat}, {Saintonge}, {Brosch}, {Catinella}, {Hoffman}, {Stierwalt},
  {Spekkens}, {Lerner}, {Masters}, {Momjian}, {Rosenberg}, {Springob},
  {Boselli}, {Charmandaris}, {Darling}, {Davies}, {Garcia Lambas}, {Gavazzi},
  {Giovanardi}, {Hardy}, {Hunt}, {Iovino}, {Karachentsev}, {Karachentseva},
  {Koopmann}, {Marinoni}, {Minchin}, {Muller}, {Putman}, {Pantoja}, {Salzer},
  {Scodeggio}, {Skillman}, {Solanes}, {Valotto}, {van Driel}, \& {van
  Zee}}]{Giovanelli+2005}
{Giovanelli}, R., {Haynes}, M.~P., {Kent}, B.~R., {et~al.} 2005, \aj, 130,
  2598, \dodoi{10.1086/497431}

\bibitem[{{Guo} {et~al.}(2017){Guo}, {Li}, {Zheng}, {Mo}, {Jing}, {Zu}, {Lim},
  \& {Xu}}]{Guo+2017}
{Guo}, H., {Li}, C., {Zheng}, Z., {et~al.} 2017, \apj, 846, 61,
  \dodoi{10.3847/1538-4357/aa85e7}

\bibitem[{{Guo} {et~al.}(2015){Guo}, {Zheng}, {Zehavi}, {Behroozi}, {Chuang},
  {Comparat}, {Favole}, {Gottloeber}, {Klypin}, {Prada}, {Weinberg}, \&
  {Yepes}}]{Guo+2015}
{Guo}, H., {Zheng}, Z., {Zehavi}, I., {et~al.} 2015, \mnras, 453, 4368,
  \dodoi{10.1093/mnras/stv1966}

\bibitem[{{Haynes} {et~al.}(2011){Haynes}, {Giovanelli}, {Martin}, {Hess},
  {Saintonge}, {Adams}, {Hallenbeck}, {Hoffman}, {Huang}, {Kent}, {Koopmann},
  {Papastergis}, {Stierwalt}, {Balonek}, {Craig}, {Higdon}, {Kornreich},
  {Miller}, {O'Donoghue}, {Olowin}, {Rosenberg}, {Spekkens}, {Troischt}, \&
  {Wilcots}}]{Haynes+2011}
{Haynes}, M.~P., {Giovanelli}, R., {Martin}, A.~M., {et~al.} 2011, \aj, 142,
  170, \dodoi{10.1088/0004-6256/142/5/170}

\bibitem[{{Haynes} {et~al.}(2018){Haynes}, {Giovanelli}, {Kent}, {Adams},
  {Balonek}, {Craig}, {Fertig}, {Finn}, {Giovanardi}, {Hallenbeck}, {Hess},
  {Hoffman}, {Huang}, {Jones}, {Koopmann}, {Kornreich}, {Leisman}, {Miller},
  {Moorman}, {O'Connor}, {O'Donoghue}, {Papastergis}, {Troischt}, {Stark}, \&
  {Xiao}}]{Haynes+2018}
{Haynes}, M.~P., {Giovanelli}, R., {Kent}, B.~R., {et~al.} 2018, \apj, 861, 49,
  \dodoi{10.3847/1538-4357/aac956}

\bibitem[{{Helfer} {et~al.}(2003){Helfer}, {Thornley}, {Regan}, {Wong},
  {Sheth}, {Vogel}, {Blitz}, \& {Bock}}]{Helfer+2003}
{Helfer}, T.~T., {Thornley}, M.~D., {Regan}, M.~W., {et~al.} 2003, \apjs, 145,
  259, \dodoi{10.1086/346076}

\bibitem[{{Hu} {et~al.}(2019){Hu}, {Hoppmann}, {Staveley-Smith}, {Ger{\'e}b},
  {Oosterloo}, {Morganti}, {Catinella}, {Cortese}, {Lagos}, \&
  {Meyer}}]{Hu+2019}
{Hu}, W., {Hoppmann}, L., {Staveley-Smith}, L., {et~al.} 2019, \mnras, 489,
  1619, \dodoi{10.1093/mnras/stz2038}

\bibitem[{{Huang} {et~al.}(2012){Huang}, {Haynes}, {Giovanelli}, \&
  {Brinchmann}}]{Huang+2012}
{Huang}, S., {Haynes}, M.~P., {Giovanelli}, R., \& {Brinchmann}, J. 2012, \apj,
  756, 113, \dodoi{10.1088/0004-637X/756/2/113}

\bibitem[{{Jing} {et~al.}(2007){Jing}, {Suto}, \& {Mo}}]{Jing+2007}
{Jing}, Y.~P., {Suto}, Y., \& {Mo}, H.~J. 2007, \apj, 657, 664,
  \dodoi{10.1086/511130}

\bibitem[{{Jones} {et~al.}(2018){Jones}, {Haynes}, {Giovanelli}, \&
  {Moorman}}]{Jones+2018}
{Jones}, M.~G., {Haynes}, M.~P., {Giovanelli}, R., \& {Moorman}, C. 2018,
  \mnras, 477, 2, \dodoi{10.1093/mnras/sty521}

\bibitem[{{Jones} {et~al.}(2020){Jones}, {Hess}, {Adams}, \&
  {Verdes-Montenegro}}]{Jones+2020}
{Jones}, M.~G., {Hess}, K.~M., {Adams}, E. A.~K., \& {Verdes-Montenegro}, L.
  2020, arXiv e-prints, arXiv:2003.09302.
\newblock \doarXiv{2003.09302}

\bibitem[{{Kere{\v{s}}} {et~al.}(2009){Kere{\v{s}}}, {Katz}, {Fardal},
  {Dav{\'e}}, \& {Weinberg}}]{Keres+2009}
{Kere{\v{s}}}, D., {Katz}, N., {Fardal}, M., {Dav{\'e}}, R., \& {Weinberg},
  D.~H. 2009, \mnras, 395, 160, \dodoi{10.1111/j.1365-2966.2009.14541.x}

\bibitem[{{Kere{\v{s}}} {et~al.}(2005){Kere{\v{s}}}, {Katz}, {Weinberg}, \&
  {Dav{\'e}}}]{Keres+2005}
{Kere{\v{s}}}, D., {Katz}, N., {Weinberg}, D.~H., \& {Dav{\'e}}, R. 2005,
  \mnras, 363, 2, \dodoi{10.1111/j.1365-2966.2005.09451.x}

\bibitem[{{Kim} {et~al.}(2017){Kim}, {Wyithe}, {Baugh}, {Lagos}, {Power}, \&
  {Park}}]{Kim+2017}
{Kim}, H.-S., {Wyithe}, J.~S.~B., {Baugh}, C.~M., {et~al.} 2017, \mnras, 465,
  111, \dodoi{10.1093/mnras/stw2779}

\bibitem[{{Krumholz}(2013)}]{Krumholz+2013}
{Krumholz}, M.~R. 2013, \mnras, 436, 2747, \dodoi{10.1093/mnras/stt1780}

\bibitem[{{Lagos} {et~al.}(2017){Lagos}, {Theuns}, {Stevens}, {Cortese},
  {Padilla}, {Davis}, {Contreras}, \& {Croton}}]{Lagos+2017}
{Lagos}, C. d.~P., {Theuns}, T., {Stevens}, A. R.~H., {et~al.} 2017, \mnras,
  464, 3850, \dodoi{10.1093/mnras/stw2610}

\bibitem[{{Lagos} {et~al.}(2018){Lagos}, {Tobar}, {Robotham}, {Obreschkow},
  {Mitchell}, {Power}, \& {Elahi}}]{Lagos+2018}
{Lagos}, C. d.~P., {Tobar}, R.~J., {Robotham}, A. S.~G., {et~al.} 2018, \mnras,
  481, 3573, \dodoi{10.1093/mnras/sty2440}

\bibitem[{{Lah} {et~al.}(2007){Lah}, {Chengalur}, {Briggs}, {Colless}, {de
  Propris}, {Pracy}, {de Blok}, {Fujita}, {Ajiki}, {Shioya}, {Nagao},
  {Murayama}, {Taniguchi}, {Yagi}, \& {Okamura}}]{Lah+2007}
{Lah}, P., {Chengalur}, J.~N., {Briggs}, F.~H., {et~al.} 2007, \mnras, 376,
  1357, \dodoi{10.1111/j.1365-2966.2007.11540.x}

\bibitem[{{Leroy} {et~al.}(2009){Leroy}, {Walter}, {Bigiel}, {Usero}, {Weiss},
  {Brinks}, {de Blok}, {Kennicutt}, {Schuster}, {Kramer}, {Wiesemeyer}, \&
  {Roussel}}]{Leroy+2009}
{Leroy}, A.~K., {Walter}, F., {Bigiel}, F., {et~al.} 2009, \aj, 137, 4670,
  \dodoi{10.1088/0004-6256/137/6/4670}

\bibitem[{{Li} {et~al.}(2012){Li}, {Kauffmann}, {Fu}, {Wang}, {Catinella},
  {Fabello}, {Schiminovich}, \& {Zhang}}]{Li+2012}
{Li}, C., {Kauffmann}, G., {Fu}, J., {et~al.} 2012, \mnras, 424, 1471,
  \dodoi{10.1111/j.1365-2966.2012.21337.x}

\bibitem[{{Li} \& {White}(2009)}]{Li+2009}
{Li}, C., \& {White}, S.~D.~M. 2009, \mnras, 398, 2177,
  \dodoi{10.1111/j.1365-2966.2009.15268.x}

\bibitem[{{Lim} {et~al.}(2017){Lim}, {Mo}, {Lu}, {Wang}, \& {Yang}}]{Lim+2017}
{Lim}, S.~H., {Mo}, H.~J., {Lu}, Y., {Wang}, H., \& {Yang}, X. 2017, \mnras,
  470, 2982, \dodoi{10.1093/mnras/stx1462}

\bibitem[{{Lutz} {et~al.}(2018){Lutz}, {Kilborn}, {Koribalski}, {Catinella},
  {J{\'o}zsa}, {Wong}, {Stevens}, {Obreschkow}, \& {D{\'e}nes}}]{Lutz+2018}
{Lutz}, K.~A., {Kilborn}, V.~A., {Koribalski}, B.~S., {et~al.} 2018, \mnras,
  476, 3744, \dodoi{10.1093/mnras/sty387}

\bibitem[{{Maddox} {et~al.}(2015){Maddox}, {Hess}, {Obreschkow}, {Jarvis}, \&
  {Blyth}}]{Maddox+2015}
{Maddox}, N., {Hess}, K.~M., {Obreschkow}, D., {Jarvis}, M.~J., \& {Blyth},
  S.~L. 2015, \mnras, 447, 1610, \dodoi{10.1093/mnras/stu2532}

\bibitem[{{Man} \& {Belli}(2018)}]{Man+2018}
{Man}, A., \& {Belli}, S. 2018, Nature Astronomy, 2, 695,
  \dodoi{10.1038/s41550-018-0558-1}

\bibitem[{{Marinacci} {et~al.}(2018){Marinacci}, {Vogelsberger}, {Pakmor},
  {Torrey}, {Springel}, {Hernquist}, {Nelson}, {Weinberger}, {Pillepich},
  {Naiman}, \& {Genel}}]{Marinacci+2018}
{Marinacci}, F., {Vogelsberger}, M., {Pakmor}, R., {et~al.} 2018, \mnras, 480,
  5113, \dodoi{10.1093/mnras/sty2206}

\bibitem[{{Martin} {et~al.}(2010){Martin}, {Papastergis}, {Giovanelli},
  {Haynes}, {Springob}, \& {Stierwalt}}]{Martin+2010}
{Martin}, A.~M., {Papastergis}, E., {Giovanelli}, R., {et~al.} 2010, \apj, 723,
  1359, \dodoi{10.1088/0004-637X/723/2/1359}

\bibitem[{{Meyer} {et~al.}(2017){Meyer}, {Robotham}, {Obreschkow}, {Westmeier},
  {Duffy}, \& {Staveley-Smith}}]{Meyer+2017}
{Meyer}, M., {Robotham}, A., {Obreschkow}, D., {et~al.} 2017, \pasa, 34, 52,
  \dodoi{10.1017/pasa.2017.31}

\bibitem[{{Meyer} {et~al.}(2004){Meyer}, {Zwaan}, {Webster}, {Staveley-Smith},
  {Ryan-Weber}, {Drinkwater}, {Barnes}, {Howlett}, {Kilborn}, {Stevens},
  {Waugh}, {Pierce}, {Bhathal}, {de Blok}, {Disney}, {Ekers}, {Freeman},
  {Garcia}, {Gibson}, {Harnett}, {Henning}, {Jerjen}, {Kesteven}, {Knezek},
  {Koribalski}, {Mader}, {Marquarding}, {Minchin}, {O'Brien}, {Oosterloo},
  {Price}, {Putman}, {Ryder}, {Sadler}, {Stewart}, {Stootman}, \&
  {Wright}}]{Meyer+2004}
{Meyer}, M.~J., {Zwaan}, M.~A., {Webster}, R.~L., {et~al.} 2004, \mnras, 350,
  1195, \dodoi{10.1111/j.1365-2966.2004.07710.x}

\bibitem[{{Moster} {et~al.}(2010){Moster}, {Somerville}, {Maulbetsch}, {van den
  Bosch}, {Macci{\`o}}, {Naab}, \& {Oser}}]{Moster+2010}
{Moster}, B.~P., {Somerville}, R.~S., {Maulbetsch}, C., {et~al.} 2010, \apj,
  710, 903, \dodoi{10.1088/0004-637X/710/2/903}

\bibitem[{{Naiman} {et~al.}(2018){Naiman}, {Pillepich}, {Springel},
  {Ramirez-Ruiz}, {Torrey}, {Vogelsberger}, {Pakmor}, {Nelson}, {Marinacci},
  {Hernquist}, {Weinberger}, \& {Genel}}]{Naiman+2018}
{Naiman}, J.~P., {Pillepich}, A., {Springel}, V., {et~al.} 2018, \mnras, 477,
  1206, \dodoi{10.1093/mnras/sty618}

\bibitem[{{Nelson} {et~al.}(2018){Nelson}, {Pillepich}, {Springel},
  {Weinberger}, {Hernquist}, {Pakmor}, {Genel}, {Torrey}, {Vogelsberger},
  {Kauffmann}, {Marinacci}, \& {Naiman}}]{Nelson+2018}
{Nelson}, D., {Pillepich}, A., {Springel}, V., {et~al.} 2018, \mnras, 475, 624,
  \dodoi{10.1093/mnras/stx3040}

\bibitem[{{Obreschkow} {et~al.}(2016){Obreschkow}, {Glazebrook}, {Kilborn}, \&
  {Lutz}}]{Obreschkow+2016}
{Obreschkow}, D., {Glazebrook}, K., {Kilborn}, V., \& {Lutz}, K. 2016, \apjl,
  824, L26, \dodoi{10.3847/2041-8205/824/2/L26}

\bibitem[{{Obuljen} {et~al.}(2019){Obuljen}, {Alonso}, {Villaescusa-Navarro},
  {Yoon}, \& {Jones}}]{Obuljen+2019}
{Obuljen}, A., {Alonso}, D., {Villaescusa-Navarro}, F., {Yoon}, I., \& {Jones},
  M. 2019, \mnras, 486, 5124, \dodoi{10.1093/mnras/stz1118}

\bibitem[{{Papastergis} {et~al.}(2011){Papastergis}, {Martin}, {Giovanelli}, \&
  {Haynes}}]{Papastergis+2011}
{Papastergis}, E., {Martin}, A.~M., {Giovanelli}, R., \& {Haynes}, M.~P. 2011,
  \apj, 739, 38, \dodoi{10.1088/0004-637X/739/1/38}

\bibitem[{{Paul} {et~al.}(2018){Paul}, {Choudhury}, \& {Paranjape}}]{Paul+2018}
{Paul}, N., {Choudhury}, T.~R., \& {Paranjape}, A. 2018, \mnras, 479, 1627,
  \dodoi{10.1093/mnras/sty1539}

\bibitem[{{Pillepich} {et~al.}(2018){Pillepich}, {Nelson}, {Hernquist},
  {Springel}, {Pakmor}, {Torrey}, {Weinberger}, {Genel}, {Naiman}, {Marinacci},
  \& {Vogelsberger}}]{Pillepich+2018}
{Pillepich}, A., {Nelson}, D., {Hernquist}, L., {et~al.} 2018, \mnras, 475,
  648, \dodoi{10.1093/mnras/stx3112}

\bibitem[{{Planck Collaboration} {et~al.}(2014){Planck Collaboration}, {Ade},
  {Aghanim}, \& et~al.}]{PlanckCollaboration+2014}
{Planck Collaboration}, {Ade}, P.~A.~R., {Aghanim}, N., \& et~al. 2014, \aap,
  571, A16, \dodoi{10.1051/0004-6361/201321591}

\bibitem[{{Popping} {et~al.}(2015){Popping}, {Behroozi}, \&
  {Peeples}}]{Popping+2015}
{Popping}, G., {Behroozi}, P.~S., \& {Peeples}, M.~S. 2015, \mnras, 449, 477,
  \dodoi{10.1093/mnras/stv318}

\bibitem[{{Rees} \& {Ostriker}(1977)}]{Rees+1977}
{Rees}, M.~J., \& {Ostriker}, J.~P. 1977, \mnras, 179, 541,
  \dodoi{10.1093/mnras/179.4.541}

\bibitem[{{Rhee} {et~al.}(2013){Rhee}, {Zwaan}, {Briggs}, {Chengalur}, {Lah},
  {Oosterloo}, \& {van der Hulst}}]{Rhee+2013}
{Rhee}, J., {Zwaan}, M.~A., {Briggs}, F.~H., {et~al.} 2013, \mnras, 435, 2693,
  \dodoi{10.1093/mnras/stt1481}

\bibitem[{{Saintonge} {et~al.}(2011){Saintonge}, {Kauffmann}, {Kramer},
  {Tacconi}, {Buchbender}, {Catinella}, {Fabello}, {Graci{\'a}-Carpio}, {Wang},
  {Cortese}, {Fu}, {Genzel}, {Giovanelli}, {Guo}, {Haynes}, {Heckman},
  {Krumholz}, {Lemonias}, {Li}, {Moran}, {Rodriguez-Fernandez}, {Schiminovich},
  {Schuster}, \& {Sievers}}]{Saintonge+2011}
{Saintonge}, A., {Kauffmann}, G., {Kramer}, C., {et~al.} 2011, \mnras, 415, 32,
  \dodoi{10.1111/j.1365-2966.2011.18677.x}

\bibitem[{{Saintonge} {et~al.}(2018){Saintonge}, {Wilson}, {Xiao}, {Lin},
  {Hwang}, {Tosaki}, {Bureau}, {Cigan}, {Clark}, {Clements}, {De Looze},
  {Dharmawardena}, {Gao}, {Gear}, {Greenslade}, {Lamperti}, {Lee}, {Li},
  {Micha{\l}owski}, {Mok}, {Pan}, {Sansom}, {Sargent}, {Smith}, {Williams},
  {Yang}, {Zhu}, {Accurso}, {Barmby}, {Brinks}, {Bourne}, {Brown}, {Chung},
  {Chung}, {Cibinel}, {Coppin}, {Davies}, {Davis}, {Eales}, {Fanciullo},
  {Fang}, {Gao}, {Glass}, {Gomez}, {Greve}, {He}, {Ho}, {Huang}, {Jeong},
  {Jiang}, {Jiao}, {Kemper}, {Kim}, {Kim}, {Kim}, {Ko}, {Kong}, {Lacaille},
  {Lacey}, {Lee}, {Lee}, {Lee}, {Masters}, {Oh}, {Papadopoulos}, {Park},
  {Park}, {Parsons}, {Rowland s}, {Scicluna}, {Scudder}, {Sethuram},
  {Serjeant}, {Shao}, {Sheen}, {Shi}, {Shim}, {Smith}, {Spekkens}, {Tsai},
  {Verma}, {Urquhart}, {Violino}, {Viti}, {Wake}, {Wang}, {Wouterloot}, {Yang},
  {Yim}, {Yuan}, \& {Zheng}}]{Saintonge+2018}
{Saintonge}, A., {Wilson}, C.~D., {Xiao}, T., {et~al.} 2018, \mnras, 481, 3497,
  \dodoi{10.1093/mnras/sty2499}

\bibitem[{{Sato-Polito} {et~al.}(2019){Sato-Polito}, {Montero-Dorta}, {Abramo},
  {Prada}, \& {Klypin}}]{Sato-Polito+2019}
{Sato-Polito}, G., {Montero-Dorta}, A.~D., {Abramo}, L.~R., {Prada}, F., \&
  {Klypin}, A. 2019, \mnras, 487, 1570, \dodoi{10.1093/mnras/stz1338}

\bibitem[{{Silk}(1977)}]{Silk+1977}
{Silk}, J. 1977, \apj, 211, 638, \dodoi{10.1086/154972}

\bibitem[{{Springel} {et~al.}(2018){Springel}, {Pakmor}, {Pillepich},
  {Weinberger}, {Nelson}, {Hernquist}, {Vogelsberger}, {Genel}, {Torrey},
  {Marinacci}, \& {Naiman}}]{Springel+2018}
{Springel}, V., {Pakmor}, R., {Pillepich}, A., {et~al.} 2018, \mnras, 475, 676,
  \dodoi{10.1093/mnras/stx3304}

\bibitem[{{Stevens} {et~al.}(2019){Stevens}, {Diemer}, {Lagos}, {Nelson},
  {Pillepich}, {Brown}, {Catinella}, {Hernquist}, {Weinberger}, {Vogelsberger},
  \& {Marinacci}}]{Stevens+2019}
{Stevens}, A. R.~H., {Diemer}, B., {Lagos}, C. d.~P., {et~al.} 2019, \mnras,
  483, 5334, \dodoi{10.1093/mnras/sty3451}

\bibitem[{{van de Voort} {et~al.}(2011){van de Voort}, {Schaye}, {Booth},
  {Haas}, \& {Dalla Vecchia}}]{vandeVoort+2011}
{van de Voort}, F., {Schaye}, J., {Booth}, C.~M., {Haas}, M.~R., \& {Dalla
  Vecchia}, C. 2011, \mnras, 414, 2458,
  \dodoi{10.1111/j.1365-2966.2011.18565.x}

\bibitem[{{Verheijen} {et~al.}(2007){Verheijen}, {van Gorkom}, {Szomoru},
  {Dwarakanath}, {Poggianti}, \& {Schiminovich}}]{Verheijen+2007}
{Verheijen}, M., {van Gorkom}, J.~H., {Szomoru}, A., {et~al.} 2007, \apjl, 668,
  L9, \dodoi{10.1086/522621}

\bibitem[{{Villaescusa-Navarro} {et~al.}(2018){Villaescusa-Navarro}, {Genel},
  {Castorina}, {Obuljen}, {Spergel}, {Hernquist}, {Nelson}, {Carucci},
  {Pillepich}, {Marinacci}, {Diemer}, {Vogelsberger}, {Weinberger}, \&
  {Pakmor}}]{Villaescusa-Navarro+2018}
{Villaescusa-Navarro}, F., {Genel}, S., {Castorina}, E., {et~al.} 2018, \apj,
  866, 135, \dodoi{10.3847/1538-4357/aadba0}

\bibitem[{{Vogelsberger} {et~al.}(2014){Vogelsberger}, {Genel}, {Springel},
  {Torrey}, {Sijacki}, {Xu}, {Snyder}, {Nelson}, \&
  {Hernquist}}]{Vogelsberger+2014}
{Vogelsberger}, M., {Genel}, S., {Springel}, V., {et~al.} 2014, \mnras, 444,
  1518, \dodoi{10.1093/mnras/stu1536}

\bibitem[{{Wang} {et~al.}(2016){Wang}, {Koribalski}, {Serra}, {van der Hulst},
  {Roychowdhury}, {Kamphuis}, \& {Chengalur}}]{Wang+2016}
{Wang}, J., {Koribalski}, B.~S., {Serra}, P., {et~al.} 2016, \mnras, 460, 2143,
  \dodoi{10.1093/mnras/stw1099}

\bibitem[{{Wechsler} {et~al.}(2006){Wechsler}, {Zentner}, {Bullock},
  {Kravtsov}, \& {Allgood}}]{Wechsler+2006}
{Wechsler}, R.~H., {Zentner}, A.~R., {Bullock}, J.~S., {Kravtsov}, A.~V., \&
  {Allgood}, B. 2006, \apj, 652, 71, \dodoi{10.1086/507120}

\bibitem[{{White} \& {Rees}(1978)}]{White+1978}
{White}, S.~D.~M., \& {Rees}, M.~J. 1978, \mnras, 183, 341,
  \dodoi{10.1093/mnras/183.3.341}

\bibitem[{{Wolz} {et~al.}(2019){Wolz}, {Murray}, {Blake}, \&
  {Wyithe}}]{Wolz+2019}
{Wolz}, L., {Murray}, S.~G., {Blake}, C., \& {Wyithe}, J.~S. 2019, \mnras, 484,
  1007, \dodoi{10.1093/mnras/sty3142}

\bibitem[{{Xu} \& {Zheng}(2018)}]{Xu+2018}
{Xu}, X., \& {Zheng}, Z. 2018, \mnras, 479, 1579, \dodoi{10.1093/mnras/sty1547}

\bibitem[{{Yang} {et~al.}(2007){Yang}, {Mo}, {van den Bosch}, {Pasquali}, {Li},
  \& {Barden}}]{Yang+2007}
{Yang}, X., {Mo}, H.~J., {van den Bosch}, F.~C., {et~al.} 2007, \apj, 671, 153,
  \dodoi{10.1086/522027}

\bibitem[{{Yang} {et~al.}(2012){Yang}, {Mo}, {van den Bosch}, {Zhang}, \&
  {Han}}]{Yang+2012}
{Yang}, X., {Mo}, H.~J., {van den Bosch}, F.~C., {Zhang}, Y., \& {Han}, J.
  2012, \apj, 752, 41, \dodoi{10.1088/0004-637X/752/1/41}

\bibitem[{{York} {et~al.}(2000){York}, {Adelman}, {Anderson}, {Anderson},
  {Annis}, {Bahcall}, {Bakken}, {Barkhouser}, {Bastian}, {Berman}, {Boroski},
  {Bracker}, {Briegel}, {Briggs}, {Brinkmann}, {Brunner}, {Burles}, {Carey},
  {Carr}, {Castander}, {Chen}, {Colestock}, {Connolly}, {Crocker}, {Csabai},
  {Czarapata}, {Davis}, {Doi}, {Dombeck}, {Eisenstein}, {Ellman}, {Elms},
  {Evans}, {Fan}, {Federwitz}, {Fiscelli}, {Friedman}, {Frieman}, {Fukugita},
  {Gillespie}, {Gunn}, {Gurbani}, {de Haas}, {Haldeman}, {Harris}, {Hayes},
  {Heckman}, {Hennessy}, {Hindsley}, {Holm}, {Holmgren}, {Huang}, {Hull},
  {Husby}, {Ichikawa}, {Ichikawa}, {Ivezi{\'c}}, {Kent}, {Kim}, {Kinney},
  {Klaene}, {Kleinman}, {Kleinman}, {Knapp}, {Korienek}, {Kron}, {Kunszt},
  {Lamb}, {Lee}, {Leger}, {Limmongkol}, {Lindenmeyer}, {Long}, {Loomis},
  {Loveday}, {Lucinio}, {Lupton}, {MacKinnon}, {Mannery}, {Mantsch}, {Margon},
  {McGehee}, {McKay}, {Meiksin}, {Merelli}, {Monet}, {Munn}, {Narayanan},
  {Nash}, {Neilsen}, {Neswold}, {Newberg}, {Nichol}, {Nicinski}, {Nonino},
  {Okada}, {Okamura}, {Ostriker}, {Owen}, {Pauls}, {Peoples}, {Peterson},
  {Petravick}, {Pier}, {Pope}, {Pordes}, {Prosapio}, {Rechenmacher}, {Quinn},
  {Richards}, {Richmond}, {Rivetta}, {Rockosi}, {Ruthmansdorfer}, {Sandford},
  {Schlegel}, {Schneider}, {Sekiguchi}, {Sergey}, {Shimasaku}, {Siegmund},
  {Smee}, {Smith}, {Snedden}, {Stone}, {Stoughton}, {Strauss}, {Stubbs},
  {SubbaRao}, {Szalay}, {Szapudi}, {Szokoly}, {Thakar}, {Tremonti}, {Tucker},
  {Uomoto}, {Vanden Berk}, {Vogeley}, {Waddell}, {Wang}, {Watanabe},
  {Weinberg}, {Yanny}, {Yasuda}, \& {SDSS Collaboration}}]{York+2000}
{York}, D.~G., {Adelman}, J., {Anderson}, Jr., J.~E., {et~al.} 2000, \aj, 120,
  1579, \dodoi{10.1086/301513}

\bibitem[{{Zhang} {et~al.}(2009){Zhang}, {Li}, {Kauffmann}, {Zou}, {Catinella},
  {Shen}, {Guo}, \& {Chang}}]{Zhang+2009}
{Zhang}, W., {Li}, C., {Kauffmann}, G., {et~al.} 2009, \mnras, 397, 1243,
  \dodoi{10.1111/j.1365-2966.2009.15050.x}

\bibitem[{{Zoldan} {et~al.}(2017){Zoldan}, {De Lucia}, {Xie}, {Fontanot}, \&
  {Hirschmann}}]{Zoldan+2017}
{Zoldan}, A., {De Lucia}, G., {Xie}, L., {Fontanot}, F., \& {Hirschmann}, M.
  2017, \mnras, 465, 2236, \dodoi{10.1093/mnras/stw2901}

\end{thebibliography}

\end{document}